\documentclass[prd, 12pt, superscriptaddress, nofootinbib,preprintnumbers,
reprint,
onecolumn,
floatfix
]{revtex4-2} 
\usepackage{amsmath,amssymb}
\usepackage{mathtools}

\usepackage{graphicx}
\usepackage{dcolumn}
\usepackage{bm}
\usepackage{natbib}
\usepackage{url}
\usepackage{tikz}
\usetikzlibrary{shapes}
\usetikzlibrary{positioning}
\usetikzlibrary{calc} 
\usepackage{qcircuit}
\usepackage{braket}
\usepackage{bbm}
\usepackage{esvect}
\usepackage{ulem}
\usepackage[colorlinks=true,backref=false, linktocpage=true,
citecolor=blue,urlcolor=blue,linkcolor=blue,pdfpagemode=UseOutlines]{hyperref}
\usepackage{multirow}
\usepackage[section]{placeins}
\usepackage{booktabs}
\usepackage{appendix}
\usepackage{empheq}
\usepackage{array}
\usepackage{subcaption}
\usepackage{todonotes}

\tikzset{
  thickdot/.style={circle, fill=black, minimum size=6pt, inner sep=0pt, draw},
  connectline/.style={line width=0.8pt}
}

\begin{document}


\title{Quantum Utility in Simulating the Real-time Dynamics of the Fermi-Hubbard Model using Superconducting Quantum Computers}

 \newcommand*{\KU}{Department of Physics and Astronomy, University of Kansas, Lawrence, Kansas 66045, USA.}\affiliation{\KU}
\newcommand*{\SBU}{C. N. Yang Institute for Theoretical Physics,
Stony Brook University, New York 11794, USA.}\affiliation{\SBU}
\newcommand*{\IBM}{IBM Quantum, IBM Thomas J. Watson Research Center, Yorktown Heights, NY 10598, USA.}\affiliation{\IBM}
\newcommand*{\CSIBNL}{Computational Science Department, Brookhaven National Laboratory, Upton, New York 11973, USA.}\affiliation{\CSIBNL}

\author{Talal Ahmed Chowdhury}\email[Correspondence to: ] {talal@ku.edu}\affiliation{\KU}
\author{Vladimir Korepin}\email{vladimir.korepin@stonybrook.edu}\affiliation{\SBU}
\author{Vincent R. Pascuzzi}\email{vrpascuzzi@ibm.com}\affiliation{\IBM}
\author{Kwangmin Yu}\email[Correspondence to: ] {kyu@bnl.gov}\affiliation{\CSIBNL}

\begin{abstract}
The Fermi-Hubbard model is a fundamental model in condensed matter physics that describes strongly correlated electrons. On the other hand, quantum computers are emerging as powerful tools for exploring the complex dynamics of these quantum many-body systems. In this work, we demonstrate the quantum simulation of the one-dimensional Fermi-Hubbard model using IBM's superconducting quantum computers, employing over 100 qubits. We introduce a first-order Trotterization scheme and extend it to an optimized second-order Trotterization for the time evolution in the Fermi-Hubbard model, specifically tailored for the limited qubit connectivity of quantum architectures, such as IBM's platforms. Notably, both Trotterization approaches are scalable and maintain a constant circuit depth at each Trotter step, regardless of the qubit count, enabling us to precisely investigate the relaxation dynamics in the Fermi-Hubbard model by measuring the expectation value of the Néel observable (staggered magnetization) for time-evolved quantum states. Finally, our successful measurement of expectation values in such large-scale quantum many-body systems, especially at longer time scales with larger entanglement, highlights the quantum utility of superconducting quantum platforms over conventional classical approximation methods.
\end{abstract}

\maketitle

\section{Introduction}\label{sec:introduction}
Quantum computers are anticipated to play a crucial role in exploring the complex dynamics of many-body quantum systems. Therefore, validating existing results on current quantum computers is essential for assessing their potential future applications. Real quantum devices are emerging as effective programmable quantum platforms, capable of accurately preparing intricate quantum many-body models. These platforms offer a valuable testing ground for investigating novel phenomena that would be challenging or impossible to realize through experimental methods~\cite{Swan-dynamics-quantum-info-review, Georgescu:2013oza, Daley:2022eja}.

In this study, we focus on the large-scale quantum simulation of the Fermi-Hubbard model using IBM's superconducting quantum computers, specifically examining its real-time dynamics and utilizing these quantum devices as programmable platforms. The Fermi-Hubbard model~\cite{Hubbard-1} is a fundamental representation of itinerant, interacting electrons on various lattice geometries and dimensionalities that captures many properties and phenomena found in strongly correlated systems, such as the electronic behavior of solids with narrow bands, magnetism in ferromagnetic transition metals, the Mott metal-insulator transition, and the electronic properties of high-temperature cuprate superconductors in their normal state. Despite its seemingly simple structure, the Fermi-Hubbard model lacks analytical solutions in general dimensions. However, the one-dimensional Fermi-Hubbard model~\cite{Essler-Frahm-Gohmann-Klumper-Korepin-2005, Deguchi}~(and references therein) has a distinctive feature of being an integrable model and, importantly, serves not only as a model to understand strongly correlated fermionic systems in one dimension but also provides the benchmark for the development of approximate and numerical methods. Moreover, the one-dimensional Fermi-Hubbard model displays a phenomenon known as the spin-charge separation, where the spin and charge excitations propagate at different speeds due to the interparticle interactions, a feature absent in higher dimensions~\cite{Coll, Haldane-Luttinger, Masao-Shiba, Schulz-correlated, Kim-spin-charge-separation, Segovia-spin-charge-separation, Kim-spinon-holon}. Additionally, as an integrable model, the one-dimensional Fermi-Hubbard model possesses an extensive set of conserved quantities that demonstrate how non-thermal quantum quench dynamics lead to anomalous transport properties~\cite{Kinoshita, Rigol-relaxation, Vidmar-Rigol, Gring-relaxation, Langen-generalized, Vasseur-Moore}. 

To study the real-time dynamics in a one-dimensional Fermi-Hubbard model, we present a first-order Trotterization implementation, as well as qubit encoding, of the corresponding time evolution operator designed to be well-suited for quantum computers having limited connections between qubits, such as superconducting quantum computing platforms. Furthermore, we extend the first-order implementation to an optimized second-order Trotterization. Based on the new implementations,
we highlight the quantum utility of IBM's quantum computers, demonstrated previously for quantum many-body systems with more than 100 qubits~\cite{kim2023evidence,chowdhury2024enhancing, choi2025quantum}, by investigating the quantum quench dynamics through the large-scale quantum simulation of the one-dimensional Fermi-Hubbard model. Quantum simulation algorithms for the Fermi-Hubbard model have been studied extensively in~\cite{Ortiz-Gubernatis-Knill-Laflamme, Wecker-Hastings-Weibe-Clark-Nayak-Troyer, Babbush-Low-depth, Kivlichan-QS-electronic,Jiang-Sung-Kechedzhi-Smelyanski-Boixo, Kivlichan-Improved, Campbell-FHM}, and on the real quantum computers in~\cite{Hubbard-google, Hartree-Fock-google, spin-charge-google, Vilchez-Estevez:2025zjm}. Here, we specifically focus on the time evolution of the N\'eel state, which is a product state where single spin-up electrons and single spin-down electrons occupy the even and odd sites of the chain, respectively, and is not an eigenstate of the Fermi-Hubbard Hamiltonian, and capture the relaxation dynamics by determining the staggered magnetization during the time evolution. Our approach ensures that the total circuit depth of the time-evolution quantum circuit remains constant even as we increase the number of fermion sites $L$, which corresponds to the number of qubits, $N=2L$. As a result, our Trotterization method is scalable and capable of handling large-scale quantum simulations, with limitations imposed by the coherence time and noises of the devices themselves. We therefore employ a combination of quantum error mitigation techniques, including Twirled Readout Error Extinction (TREX), dynamical decoupling (DD), Pauli twirling (PT), and zero-noise extrapolation (ZNE), to enhance the precision of our observable measurements in large-scale simulations on real quantum devices. As a result, we achieve the time evolution of the staggered magnetization for fermionic system sizes involving more than 100 qubits of IBM quantum computers, which are beyond the capability of classical exact methods.

The structure of this paper is organized as follows. In section~ \ref{sec:model}, we provide a concise introduction to the one-dimensional Fermi-Hubbard model and its mapping onto qubits. Section \ref{sec:trotterization} presents our first-order and optimized second-order Trotter circuits for the time evolution operator in the one-dimensional Fermi-Hubbard model. In section~\ref{sec:implementation}, we describe the implementation of the Trotter circuits on IBM quantum computers. Section~\ref{sec:discussion} discusses the results of the variation in the expectation value of the N\'eel observable during the time evolution driven by the Fermi-Hubbard Hamiltonian. Finally, we draw our conclusions in section~\ref{sec:conclusion}. In Appendix~\ref{app:error_mitigations}, we provide a detailed discussion of the quantum error mitigation methods used in our study.

\section{The Fermi-Hubbard Model}\label{sec:model}

The Fermi-Hubbard model on $L$ lattice sites in one dimension with the open boundary conditions is given by,
\begin{equation}
	H = -t \sum_{j=0}^{L-2}\sum_{a = \uparrow, \downarrow} (c^{\dagger}_{j,a}c_{j+1,a}+c^{\dagger}_{j+1,a}c_{j,a})+U\sum_{j=0}^{L-1}n_{j,\uparrow}n_{j,\downarrow}+\sum_{j=0}^{L-1}\sum_{a=\uparrow,\downarrow}\mu_{j,a}n_{j,a}\,,
	\label{eq:FHM-model}
\end{equation}
where $c^{\dagger}_{j,a}$ and $c_{j,a}$ are the fermionic creation and annihilation operators associated to the site number $j$ and spin state $a$, and $n_{j,a}=c^{\dagger}_{j,a}c_{j,a}$ are the corresponding number operators. Besides, the total number of sites $L$ is taken as an even integer. The hopping term with coupling $t$ in Eq.~(\ref{eq:FHM-model}) describes fermions with specific spin states tunneling between the nearest-neighboring sites. The on-site interaction term with coupling $U$ introduces a repulsion for $U>0$ between two electrons in opposite spin states at the site $i$. Finally, the terms with couplings $\mu_{j,a}$ represent the local chemical potential associated with electrons at site $j$ and spin state $a$. Besides, when we set $\mu_{j,\uparrow} = \mu_{\uparrow}$ and $\mu_{j,\downarrow} = \mu_{\downarrow}$ for all sites, the Fermi-Hubbard model maintains integrability.
The operators $c^{\dagger}_{j,a}$ and $c_{j,a}$ satisfy the anticommutation relations,
\begin{equation}
	\{c_{j,a},c_{k,b}\}=\{c^{\dagger}_{j,a},c^{\dagger}_{k,b}\}=0,\,\,\,\{c_{j,a},c^{\dagger}_{k,b}\}=\delta_{jk}\delta_{ab}\,,
	\label{eq:anticomm}
\end{equation}
for $j,\,k=0,...,L-1$ and $a,\,b=\uparrow,\downarrow$. The empty lattice site $|0\rangle$ is defined by $c_{j,a}|0\rangle=0$ for all $j$ and $a$, and there are four states, $|0\rangle$, $c^{\dagger}_{j,\uparrow}|0\rangle$, $c^{\dagger}_{j,\downarrow}|0\rangle$ and $c^{\dagger}_{j,\uparrow}c^{\dagger}_{j,\downarrow}|0\rangle$, associated with every lattice site $j$. Therefore, the dimension of the corresponding Hilbert space for $L$ lattice sites is $4^{L}$.

The qubit layout of quantum devices with a heavy-hexagonal lattice topology only allows for nearest-neighbor connectivity among the qubits. Therefore, we aim to encode fermionic states in a way that ensures hopping between neighboring lattice sites or interactions between spin-up and spin-down electrons at a lattice site do not require quantum gate operations on qubits that are far apart, which is critical for large systems, as such distant operations would incur significant swap gate overhead. To achieve this, we associate the fermionic lattice site $j$ with two neighboring qubits positioned at $2j$ and $2j+1$ in the qubit register, resulting in a total of $N = 2L$ qubits. The fermionic states that span the local Hilbert space at site $j$ can be mapped into the computational basis states of the qubits $|q_{2j}, q_{2j+1}\rangle$ as follows,
\begin{equation}
|0\rangle \rightarrow |0,0\rangle,\,\,\,
c^{\dagger}_{j,\uparrow}|0\rangle \rightarrow |1,0\rangle,\,\,\,
c^{\dagger}_{j,\downarrow}|0\rangle \rightarrow |0,1\rangle,\,\,\,
c^{\dagger}_{j,\uparrow}c^{\dagger}_{j,\downarrow}|0\rangle \rightarrow |1,1\rangle\,.
\label{eq:qubit-encoding}
\end{equation}
One immediate advantage of this qubit encoding is that the terms for the chemical potential, interactions, and hopping in the Fermi-Hubbard Hamiltonian involve only single-qubit operations, nearest-neighbor qubit couplings, and next-nearest-neighbor qubit couplings, respectively, with no couplings extending beyond this range when we employ the spin-sector separated Jordan-Wigner transformation~\cite{jordan-wigner} to map the fermionic operators on qubit operators. 

Specifically, the fermionic operators at lattice site $j$ are mapped on qubit operators as,
\begin{eqnarray}
	c_{j,\uparrow}&\mapsto& Z_{0}\otimes Z_{2}...\otimes Z_{2j-2} \otimes(X_{2j}+i Y_{2j})/2\,,\nonumber\\
	c_{j,\downarrow}&\mapsto& P_{\uparrow}.Z_{1}\otimes Z_{3}...\otimes Z_{2j-1}\otimes (X_{2j+1}+i Y_{2j+1})/2\,,\label{eq:JW-map}
\end{eqnarray}
where $X_{k},\,Y_{k},\,Z_{k}$ are the Pauli matrices acting on the $k$-th qubit. Here $P_{\uparrow}$ is the global parity factor defined as,
\begin{equation}
	P_{\uparrow}=\prod_{j=0}^{L-1}Z_{2j}\,.
	\label{eq:global-parity}
\end{equation}
Under the action of the map Eq.~(\ref{eq:JW-map}), the Hamiltonian in Eq.~(\ref{eq:FHM-model}) of the 1D Fermi-Hubbard model acting on $2L$ qubits becomes,
\begin{eqnarray}
	\tilde{H} &=& -\frac{t}{2}\sum_{j=0}^{L-2}(X_{2j}X_{2j+2}+Y_{2j}Y_{2j+2}+X_{2j+1}X_{2j+3}+Y_{2j+1}Y_{2j+3})+\frac{U}{4}\sum_{j=0}^{L-1}(I_{2j}I_{2j+1}+Z_{2j} Z_{2j+1}-Z_{2j}-Z_{2j+1})\nonumber\\
    &+&\frac{1}{2}\sum_{j=0}^{L-1}\left[\mu_{j,\uparrow}(I-Z)_{2j}+\mu_{j,\downarrow}(I-Z)_{2j+1}\right]\label{eq:FHM-qubit}\,.
\end{eqnarray}
The total particle number operator, $\hat{N}_{\mathrm{tot}}$, in terms of the fermionic operators and the associated qubit operator $\tilde{N}_{\mathrm{tot}}$ are given by,
\begin{equation}
	\hat{N}_{\mathrm{tot}}=\sum_{j=0}^{L-1}(n_{j,\uparrow}+n_{j,\downarrow})\mapsto \tilde{N}_{\mathrm{tot}}=\frac{1}{2}\sum_{k=0}^{2L-1}(I-Z)_{k}\,.
	\label{eq:charge-operator}
\end{equation}
The $z$-component of the total spin operator, $\hat{S}^{z}_{\mathrm{tot}}$ in terms of fermionic operator and its associated qubit operator $\tilde{S}^{z}_{\mathrm{tot}}$ are given as,
\begin{equation}
	\hat{S}^{z}_{\mathrm{tot}}=\frac{1}{2}\sum_{j=0}^{L-1}(n_{j,\uparrow}-n_{j,\downarrow})\mapsto \tilde{S}^{z}_{\mathrm{tot}} = \frac{1}{4}\sum_{j=0}^{L-1}(Z_{2j+1}-Z_{2j})\,.
	\label{eq:magnetization}
\end{equation}
The total number of particles and the $z$-component of total spin are always conserved for the 1D Fermi-Hubbard model due to
\begin{equation}
	[H,\hat{N}_{\mathrm{tot}}]=[H,\hat{S}^{z}_{\mathrm{tot}}]=0\,.
	\label{eq:conservation}
\end{equation}
Additionally, in the limit of $\mu_{j,a}=0$ for all sites $j$ and spin-state $a$, there is an extended set of conserved charges~\cite{shastry-hubbard}. The $x$ and $y$ component of the spin operator at a site $j$ on the chain is defined as,
\begin{equation}
	\hat{S}^{x}_{j}=\frac{1}{2}(c^{\dagger}_{j,\uparrow}c_{j,\downarrow}+c^{\dagger}_{j,\downarrow}c_{j,\uparrow}),\,\,\,
	\hat{S}^{y}_{j} = \frac{1}{2i}(c^{\dagger}_{j,\uparrow}c_{j,\downarrow}-c^{\dagger}_{j,\downarrow}c_{j,\uparrow})\,,
	\label{eq:Sx-Sy}
\end{equation}
which can be mapped onto the spin operators acting on the qubits,
\begin{eqnarray}
	\tilde{S}^{x}_{j}&=&\frac{1}{4}\left[\left(\prod_{l<j}Z_{2l+1}\right)(X_{2j}X_{2j+1}+Y_{2j}Y_{2j+1})\left(\prod_{l>j}Z_{2l}\right)\right]\,,\nonumber\\
	\tilde{S}^{y}_{j}& = & \frac{1}{4}\left[\left(\prod_{l<j}Z_{2l+1}\right)(X_{2j}Y_{2j+1}-Y_{2j}X_{2j+1})\left(\prod_{l>j}Z_{2l}\right)\right]\,,\label{eq:Sx-Sy-qubit}
\end{eqnarray}
whereas the $z$-component of the spin operator $\hat{S}^{z}_{j}$ at the site $j$ and its associated operator acting on the qubits $\tilde{S}^{z}_{j}$ can be readily obtained from Eq. (\ref{eq:magnetization}).

In this work we study the N\'eel observable, $\hat{O}_{\text{N\'eel}}$, defined in terms of fermionic operators and acting on $L$ sites,
\begin{equation}
	\hat{O}_{\text{N\'eel}}=\frac{1}{2L}\sum_{j=0}^{L-1}(-1)^{j}(n_{j,\uparrow}-n_{j,\downarrow})\mapsto \tilde{O}_{\text{N\'eel}}=\frac{1}{4L}\sum_{j=0}^{L-1}(-1)^{j}(Z_{2j+1}-Z_{2j})\,,
	\label{eq:Neel-observable}
\end{equation}
where $\tilde{O}_{\text{N\'eel}}$ is the corresponding operator acting on $2L$ qubits. Since the N\'eel observable has the relation $[H,\hat{O}_{\text{N\'eel}}]\neq 0$, it will have non-trivial time evolution under the Fermi-Hubbard Hamiltonian. In particular, we determine the expectation value of the N\'eel observable, $\langle\psi(\tau)|\hat{O}_{\text{N\'eel}}|\psi(\tau)\rangle$, with respect the time-evolved state, $|\psi(\tau)\rangle=e^{-iH\tau/\hbar}|\psi_{\text{N\'eel}}\rangle$. Here, the initial N\'eel state is given as,
\begin{equation}
|\psi_{\text{N\'eel}}\rangle=c^{\dagger}_{0,\uparrow}|0\rangle\otimes c^{\dagger}_{1,\downarrow}|0\rangle\otimes c^{\dagger}_{2,\uparrow}|0\rangle\otimes c^{\dagger}_{3,\downarrow}|0\rangle\otimes...c^{\dagger}_{L-2,\uparrow}|0\rangle\otimes c^{\dagger}_{L-1,\downarrow}|0\rangle\,,
    \label{eq:neel-state}
\end{equation}
which in our qubit encoding becomes $|\psi_{\text{N\'eel}}\rangle=|10011001...1001\rangle$.

\section{Real-time Dynamics of Fermi-Hubbard model via Trotterization}\label{sec:trotterization}

The time evolution operator generated by the Fermi-Hubbard Hamiltonian is implemented for execution on a quantum computer using its Trotter-Suzuki decomposition~\cite{trotter, suzuki1, suzuki2}, commonly referred to as Trotterization. The first-order Trotterization of the time evolution operator, $U(\tau)=e^{-i \tilde{H} \tau/\hbar}$, where $\tilde{H}$ is the Jordan-Wigner-transformed Fermi-Hubbard Hamiltonian in Eq.~(\ref{eq:FHM-qubit}), acting on $2L$ qubits is given by,
\begin{equation}
	U^{(1)}(\tau)=\prod_{k=1}^{r}\hat{U}^{(1)}_{\mathrm{Trotter}_{k}}(\delta\tau),
	\label{eq:first-trotter-1}
\end{equation}
where $\delta\tau = \tau/r$ is the Trotter time-step, $r$ is the number of Trotter steps associated with the simulation time $\tau$, and $\hat{U}^{(1)}_{\mathrm{Trotter}}$ represents a single Trotter step. Now, for the Fermi-Hubbard Hamiltonian in Eq.~(\ref{eq:FHM-qubit}), denoting as $\tilde{H}=\sum_{s=1}^{M}\tilde{H}_{s}$, where $M=5L-2$ is the total number of terms in the Hamiltonian, the $\hat{U}^{(1)}_{\mathrm{Trotter}}(\delta\tau)$ is given as,
\begin{equation}
\hat{U}^{(1)}_{\mathrm{Trotter}}(\delta\tau)=\prod_{s=1}^{M}e^{-i H_{s}\delta\tau/\hbar}\,.
\label{eq:single-Trotter-1}
\end{equation}

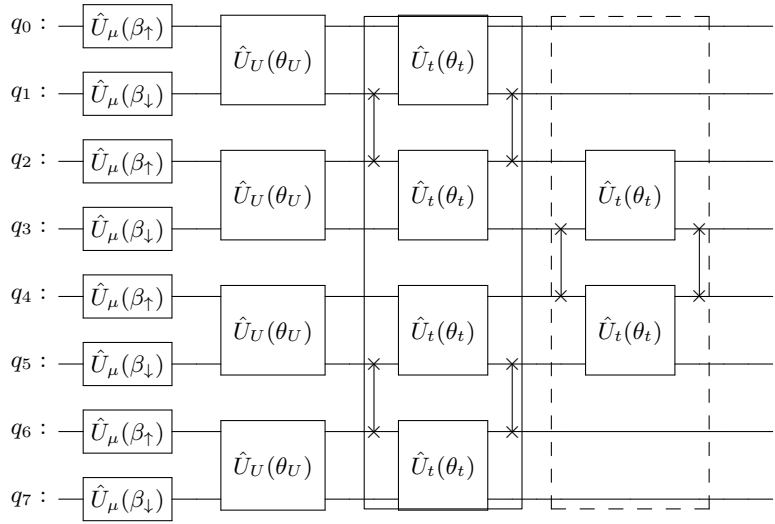
\begin{figure}[h!]
	\centerline{
		\scalebox{1.0}{
			\Qcircuit @C=1.0em @R=1.0em @!R {
				\nghost{{q}_{0} :  } & \lstick{{q}_{0} :  } & \gate{\hat{U}_{\mu}(\beta_{\uparrow})} & \qw & \multigate{1}{\hat{U}_{U}(\theta_{U})} & \qw & \qw & \multigate{1}{\hat{U}_{t}(\theta_{t})} & \qw & \qw & \qw & \qw & \qw & \qw & \qw & \qw\\
				\nghost{{q}_{1} :  } & \lstick{{q}_{1} :  } & \gate{\hat{U}_{\mu}(\beta_{\downarrow})} & \qw & \ghost{\hat{U}_{U}(\theta_{U})} & \qw & \qswap      & \ghost{\hat{U}_{t}(\theta_{t})} & \qswap      & \qw & \qw & \qw & \qw & \qw & \qw & \qw\\
				\nghost{{q}_{2} :  } & \lstick{{q}_{2} :  } & \gate{\hat{U}_{\mu}(\beta_{\uparrow})} & \qw & \multigate{1}{\hat{U}_{U}(\theta_{U})} & \qw & \qswap \qwx[-1] & \multigate{1}{\hat{U}_{t}(\theta_{t})} & \qswap \qwx[-1] & \qw & \qw & \multigate{1}{\hat{U}_{t}(\theta_{t})} & \qw & \qw & \qw & \qw\\
				\nghost{{q}_{3} :  } & \lstick{{q}_{3} :  } & \gate{\hat{U}_{\mu}(\beta_{\downarrow})} & \qw & \ghost{\hat{U}_{U}(\theta_{U})} & \qw & \qw & \ghost{\hat{U}_{t}(\theta_{t})} & \qw & \qw & \qswap & \ghost{\hat{U}_{t}(\theta_{t})} & \qswap & \qw & \qw & \qw\\
				\nghost{{q}_{4} :  } & \lstick{{q}_{4} :  } & \gate{\hat{U}_{\mu}(\beta_{\uparrow})} & \qw & \multigate{1}{\hat{U}_{U}(\theta_{U})} & \qw & \qw & \multigate{1}{\hat{U}_{t}(\theta_{t})} & \qw & \qw & \qswap \qwx[-1] & \multigate{1}{\hat{U}_{t}(\theta_{t})} & \qswap \qwx[-1] & \qw & \qw & \qw\\
				\nghost{{q}_{5} :  } & \lstick{{q}_{5} :  } & \gate{\hat{U}_{\mu}(\beta_{\downarrow})} & \qw & \ghost{\hat{U}_{U}(\theta_{U})} & \qw & \qswap & \ghost{\hat{U}_{t}(\theta_{t})} & \qswap & \qw & \qw & \ghost{\hat{U}_{t}(\theta_{t})} & \qw & \qw & \qw & \qw\\
				\nghost{{q}_{6} :  } & \lstick{{q}_{6} :  } & \gate{\hat{U}_{\mu}(\beta_{\uparrow})} & \qw & \multigate{1}{\hat{U}_{U}(\theta_{U})} & \qw & \qswap \qwx[-1] & \multigate{1}{\hat{U}_{t}(\theta_{t})} & \qswap \qwx[-1] & \qw & \qw & \qw & \qw & \qw & \qw & \qw\\
				\nghost{{q}_{7} :  } & \lstick{{q}_{7} :  } & \gate{\hat{U}_{\mu}(\beta_{\downarrow})} & \qw & \ghost{\hat{U}_{U}(\theta_{U})} & \qw & \qw & \ghost{\hat{U}_{t}(\theta_{t})} & \qw & \qw & \qw & \qw & \qw & \qw & \qw & \qw
				\gategroup{1}{7}{8}{9}{.8em}{-}
				\gategroup{1}{11}{8}{13}{.8em}{--}
			}
		}
	}
	\caption{Quantum circuit representation of the single Trotter step used in the first-order Trotterization for $L=4$ fermion sites ($N=8$ qubits).}
	\label{fig:single-trotter-step}
\end{figure}

By using the following operators,
\begin{equation}
    \hat{J}_{p,q} = X_{p}X_{q}+Y_{p}Y_{q},\,\,\,\hat{K}_{p,q}=Z_{p}Z_{q}-Z_{p}-Z_{q}\,,
    \label{eq:Pauli-ops}
\end{equation}
where, $\hat{J}_{p,q}$ and $\hat{K}_{p,q}$ operators act on $p$ and $q$-th qubits of the qubit register, the unitary operator associated with the hopping terms in Eq.~(\ref{eq:FHM-qubit}) can be written as,
\begin{equation}
    e^{i \frac{t\delta\tau}{2\hbar}\sum_{j=0}^{L-2}(\hat{J}_{2j,2j+2}+\hat{J}_{2j+1,2j+3})}\mapsto \left(\bigotimes_{\substack{j=0\\
    j(\mathrm{mod}\,2)=0}}e^{i\frac{\theta_{t}}{2}\hat{J}_{2j,2j+2}}\otimes e^{i\frac{\theta_{t}}{2}\hat{J}_{2j+1,2j+3}}\right).\left(\bigotimes_{\substack{j=0\\
    j(\mathrm{mod}\,2)=1}}e^{i\frac{\theta_{t}}{2}\hat{J}_{2j,2j+2}}\otimes e^{i\frac{\theta_{t}}{2}\hat{J}_{2j+1,2j+3}}\right)\label{eq:hopping-trotter-decompose}\,,
\end{equation}
where, we denote, $\theta_{t}=\frac{t\delta\tau}{\hbar}$. On the other hand, the unitary operator associated with the interaction terms in Eq.~(\ref{eq:FHM-qubit}) is written as,
\begin{equation}
    e^{-i\frac{U\delta\tau}{4\hbar}\sum_{j=0}^{L-1}\hat{K}_{2j,2j+1}}\mapsto \bigotimes_{j=0}^{L-1}e^{-i\frac{\theta_{U}}{4}\hat{K}_{2j,2j+1}}\,,
    \label{eq:First-Trotter-U-term}
\end{equation}
where $\theta_{U}=\frac{U\delta\tau}{\hbar}$. Finally, the unitary operator associated with the chemical potential terms in Eq.~(\ref{eq:FHM-qubit}) is written as,
\begin{equation}
    e^{i\left(\frac{\mu_{\uparrow}\delta\tau}{2\hbar}\sum_{j=0}^{L-1}Z_{2j}+\frac{\mu_{\downarrow}\delta\tau}{2\hbar}\sum_{j=0}^{L-1}Z_{2j+1}\right)}\mapsto \bigotimes_{j=0}^{L-1}\left(e^{i\frac{\beta_{\uparrow}}{2}Z_{2j}}\otimes e^{i\frac{\beta_{\downarrow}}{2}Z_{2j+1}}\right)\,,
    \label{eq:first-Trotter-chemical}
\end{equation}
where, we consider $\mu_{j,\uparrow (\downarrow)} = \mu_{\uparrow(\downarrow)}$ for sites $j$, but it can be straightforwardly generalized to site-dependent chemical potential. Besides, we deonte $\beta_{\uparrow}=\frac{\mu_{\uparrow}\delta\tau}{\hbar}$ and $\beta_{\downarrow}=\frac{\mu_{\downarrow}\delta\tau}{\hbar}$, respectively. Now, we associate the single Trotter step of the first-order Trotterization with the quantum circuit shown in Fig.~\ref{fig:single-trotter-step}, denoting it as 
\begin{equation}
	\hat{U}^{(1)}_{\mathrm{Trotter}}(\delta\tau)=\hat{\mathbf{U}}_{t}(\theta_{t}).\hat{\mathbf{U}}_{U}(\theta_{U}).\hat{\mathbf{U}}_{\mu}(\beta_{\uparrow},\beta_{\downarrow})\,,
	\label{eq:first-trotter-2}
\end{equation}
where, $\hat{\mathbf{U}}_{t}$,  $\hat{\mathbf{U}}_{U}$, and $\hat{\mathbf{U}}_{\mu}$ are the unitary operators corresponding to the circuit layers associated with the hopping terms, interaction terms and chemical potential terms described in Eq. (\ref{eq:hopping-trotter-decompose}), (\ref{eq:First-Trotter-U-term}) and (\ref{eq:first-Trotter-chemical}), respectively.

Moreover, the qubit mapping described in Eq.~(\ref{eq:qubit-encoding}) indicates that the fermionic hopping terms for the same spin states (either up-spin or down-spin) between two nearest-neighbor sites correspond to next-nearest-neighbor interactions between qubits, which is evident in the Jordan-Wigner-transformed Fermi-Hubbard Hamiltonian, Eq.~(\ref{eq:FHM-qubit}), operating on $N = 2L$ qubits. In Ref.~\cite{chowdhury2024enhancing}, efficient and scalable Trotter circuits were developed for one-dimensional Hamiltonians with such next-nearest-neighbor interactions, specifically in scenarios where qubit connectivity of a real quantum device is limited to nearest-neighbor connections. We will adopt a similar strategy by incorporating \texttt{SWAP} gates in our Trotterization implementation. While the \texttt{SWAP} gate is relatively expensive and susceptible to noise due to its decomposition into three \texttt{CNOT} gates, its effectiveness in implementation using \texttt{SWAP} gates was tested and validated in Ref. \cite{chowdhury2024enhancing} with up to 100 qubits on the IBM Heron r1 processor, \texttt{ibm\_torino}. Consequently, we present a detailed construction of the Trotterization circuits for the JW-transformed Fermi-Hubbard model, extending the approach for the qubit Hamiltonian with next-nearest-neighbor interactions.

The unitary operator associated with the hopping terms acting on the $2L$ qubits is given as,
\begin{equation}
	\hat{\mathbf{U}}_{t}(\theta_{t})=\hat{\mathbf{U}}^{t}_{\mathrm{Layer-1}}(\theta_{t}).\hat{\mathbf{U}}^{t}_{\mathrm{Layer-2}}(\theta_{t}),
	\label{eq:first-trotter-hopping}
\end{equation}
where $\hat{\mathbf{U}}^{t}_{\mathrm{Layer-1}}$ and $\hat{\mathbf{U}}^{t}_{\mathrm{Layer-2}}$ are denoted by straight-lined and dotted-lined boxes, respectively, in Fig. \ref{fig:single-trotter-step}, and are defined as,
\begin{equation}
	\hat{\mathbf{U}}^{t}_{\mathrm{Layer-1}}(\theta_{t})=\left(\bigotimes_{\substack{j=0 \\ j(\mathrm{mod}\,{2}) = 0}}^{L-1}\mathrm{SWAP}_{2j+1,2j+2}\right).\left(\bigotimes_{j=0}^{L-1}\hat{U}_{t}(\theta_{t})_{2j,2j+1}\right).\left(\bigotimes_{\substack{j=0 \\ j(\mathrm{mod}\,{2}) = 0}}^{L-1}\mathrm{SWAP}_{2j+1,2j+2}\right)\,,
	\label{eq:first-trotter-hopping-1}
\end{equation}
and
\begin{equation}
	\hat{\mathbf{U}}^{t}_{\mathrm{Layer-2}}(\theta_{t})=\left(\bigotimes_{\substack{j=0 \\ j(\mathrm{mod}\,{2}) = 1}}^{L-1}\mathrm{SWAP}_{2j+1,2j+2}\right).\left(\bigotimes_{j=1}^{L-2}\hat{U}_{t}(\theta_{t})_{2j,2j+1}\right).\left(\bigotimes_{\substack{j=0 \\ j(\mathrm{mod}\,{2}) = 1}}^{L-1}\mathrm{SWAP}_{2j+1,2j+2}\right)\,.
	\label{eq:first-trotter-hopping-2}
\end{equation}
Here, the SWAP operator acting on the qubits $(q_{2j+1}, q_{2j+2})$ and the quantum circuit of the two-qubit gate $\hat{U}_{t}(\theta_{t})$ acting on $(q_{2j},q_{2j+1})$ qubits for $j=0,..,L-1$ is given by,
\[
\vcenter{
\Qcircuit @C=1.0em @R=1.0em @!R {
     \lstick{} & \multigate{1}{\hat{U}_{t}(\theta_{t})} & \qw\\
     \lstick{} & \ghost{\hat{U}_{t}(\theta_{t})} & \qw\\
}
}
\hspace{1.0em}
=
\hspace{3.5em}
\vcenter{
\Qcircuit @C=1.0em @R=0.2em @!R {
     \lstick{{q}_{2j} :  } & \targ & \qw & \qw & \targ & \gate{R_Z(-\theta_{t})} & \targ & \gate{R_X\left(\frac{\pi}{2}\right)} & \qw\\
     \lstick{{q}_{2j+1} :  } & \ctrl{-1} & \gate{H} & \gate{R_Z\left(\theta_{t} + \frac{\pi}{2}\right)} & \ctrl{-1} & \gate{H} & \ctrl{-1} & \gate{R_X\left(-\frac{\pi}{2}\right)} & \qw\\
}
}
\]
which is based on the Yang-Baxter gate presented in Ref.~\cite{zhang2024optimal}.

The operator $\hat{\mathbf{U}}_{U}$ in the single Trotter step in Eq.~(\ref{eq:first-trotter-2}) is associated with the interaction terms of the Hamiltonian and defined as,
\begin{equation}
	\hat{\mathbf{U}}_{U}(\theta_{U})=\bigotimes_{j=0}^{L-1} \hat{U}_{U}(\theta_{U})_{2j,2j+1}\,,
	\label{eq:first-trotter-U}
\end{equation}
where the two qubit gate $\hat{U}_{U}(\theta_{U})$ acting on the qubits $(q_{2j},q_{2j+1})$ is given by
\[
\vcenter{
\Qcircuit @C=1.0em @R=1.0em @!R {
    \lstick{} & \multigate{1}{\hat{U}_{U}(\theta_{U})} & \qw\\
    \lstick{} & \ghost{\hat{U}_{U}(\theta_{U})} & \qw \\
}
}
\hspace{1.0em}
=
\hspace{3.5em}
\vcenter{
\Qcircuit @C=1.0em @R=0.2em @!R {
    \lstick{{q}_{2j} :  } & \gate{R_Z\left(\frac{\theta_{U}}{2}\right)} & \ctrl{1} & \qw & \ctrl{1} & \qw & \qw\\
    \lstick{{q}_{2j+1} :  } & \gate{R_Z\left(\frac{\theta_{U}}{2}\right)} & \targ & \gate{R_Z\left(-\frac{\theta_{U}}{2}\right)} & \targ & \qw & \qw\\
}
}
\]

The single Trotter term associated with the chemical potential terms $\hat{\mathbf{U}}_{\mu}(\beta_{\uparrow},\beta_{\downarrow})$ is given by,
\begin{equation}
	\hat{\mathbf{U}}_{\mu}(\beta_{\uparrow},\beta_{\downarrow})=\bigotimes_{j=0}^{L-1} (\hat{U}_{\mu}(\beta_{\uparrow})_{q_{2j}}\otimes \hat{U}_{\mu}(\beta_{\downarrow})_{q_{2j+1}})\,,
\end{equation}
where the single qubit operators $\hat{U}_{\mu}(\beta_{\uparrow})$ and $\hat{U}_{\mu}(\beta_{\downarrow})$ acting on $q_{2j}$ and $q_{2j+1}$ qubits, respectively, are represented by the following quantum circuit,
\[
\vcenter{
\Qcircuit @C=1.0em @R=0.5em @!R {
\lstick{q_{2j}: } & \gate{\hat{U}_{\mu}(\beta_{\uparrow})} & \qw\\
\lstick{q_{2j+1}: } & \gate{\hat{U}_{\mu}(\beta_{\downarrow})} & \qw
}
}
\hspace{1.0em}
=
\hspace{4.0em}
\vcenter{
\Qcircuit @C=1.0em @R=0.5em @!R {
\lstick{q_{2j}: } & \gate{R_{Z}(\beta_{\uparrow})} & \qw\\
\lstick{q_{2j+1}: } & \gate{R_{Z}(\beta_{\downarrow})} & \qw
}
}
\]
The main advantage of the quantum circuit implemented using first-order Trotterization presented in this work is that the depth remains constant with increasing fermion sites $L$, or the number of qubits $N=2L$, and only depends on the number of Trotter steps $r$ which can be limited by the coherence time of quantum hardware.

\begin{figure}[h!]
\centerline{
\scalebox{0.4}{
\Qcircuit @C=0.8em @R=1.2em @!R {
\lstick{q_0} & \multigate{1}{\hat{U}_U(\theta_U/2)} & \qw & \gate{\hat{U}_\mu(\beta_\uparrow/2)} & \qw & \qw & \multigate{1}{\hat{U}_t(\theta_t/2)} & \qw & \qw & \qw & \qw & \qw & \qw & \qw & \multigate{1}{\hat{U}_t(\theta_t/2)} & \qw & \qw & \gate{\hat{U}_\mu(\beta_\uparrow/2)} & \qw & \multigate{1}{\hat{U}_U(\theta_U)} & \qw & \gate{\hat{U}_\mu(\beta_\uparrow/2)} & \qw & \qw & \multigate{1}{\hat{U}_t(\theta_t/2)} & \qw & \qw & \qw & \qw & \qw & \qw & \qw & \multigate{1}{\hat{U}_t(\theta_t/2)} & \qw & \qw & \gate{\hat{U}_\mu(\beta_\uparrow/2)} & \qw & \multigate{1}{\hat{U}_U(\theta_U)} & \qw & \gate{\hat{U}_\mu(\beta_\uparrow/2)} & \qw & \qw & \multigate{1}{\hat{U}_t(\theta_t/2)} & \qw & \qw & \qw & \qw & \qw & \qw & \qw & \multigate{1}{\hat{U}_t(\theta_t/2)} & \qw & \qw & \gate{\hat{U}_\mu(\beta_\uparrow/2)} & \qw & \multigate{1}{\hat{U}_U(\theta_U/2)} & \qw & \qw & \qw \\
\lstick{q_1} & \ghost{\hat{U}_U(\theta_U/2)} & \qw & \gate{\hat{U}_\mu(\beta_\downarrow/2)} & \qw & \qswap & \ghost{\hat{U}_t(\theta_t/2)} & \qswap & \qw & \qw & \qw & \qw & \qw & \qswap & \ghost{\hat{U}_t(\theta_t/2)} & \qswap & \qw & \gate{\hat{U}_\mu(\beta_\downarrow/2)} & \qw & \ghost{\hat{U}_U(\theta_U)} & \qw & \gate{\hat{U}_\mu(\beta_\downarrow/2)} & \qw & \qswap & \ghost{\hat{U}_t(\theta_t/2)} & \qswap & \qw & \qw & \qw & \qw & \qw & \qswap & \ghost{\hat{U}_t(\theta_t/2)} & \qswap & \qw & \gate{\hat{U}_\mu(\beta_\downarrow/2)} & \qw & \ghost{\hat{U}_U(\theta_U)} & \qw & \gate{\hat{U}_\mu(\beta_\downarrow/2)} & \qw & \qswap & \ghost{\hat{U}_t(\theta_t/2)} & \qswap & \qw & \qw & \qw & \qw & \qw & \qswap & \ghost{\hat{U}_t(\theta_t/2)} & \qswap & \qw & \gate{\hat{U}_\mu(\beta_\downarrow/2)} & \qw & \ghost{\hat{U}_U(\theta_U/2)} & \qw & \qw & \qw \\
\lstick{q_2} & \multigate{1}{\hat{U}_U(\theta_U/2)} & \qw & \gate{\hat{U}_\mu(\beta_\uparrow/2)} & \qw & \qswap \qwx[-1] & \multigate{1}{\hat{U}_t(\theta_t/2)} & \qswap \qwx[-1] & \qw & \qw & \multigate{1}{\hat{U}_t(\theta_t)} & \qw & \qw & \qswap \qwx[-1] & \multigate{1}{\hat{U}_t(\theta_t/2)} & \qswap \qwx[-1] & \qw & \gate{\hat{U}_\mu(\beta_\uparrow/2)} & \qw & \multigate{1}{\hat{U}_U(\theta_U)} & \qw & \gate{\hat{U}_\mu(\beta_\uparrow/2)} & \qw & \qswap \qwx[-1] & \multigate{1}{\hat{U}_t(\theta_t/2)} & \qswap \qwx[-1] & \qw & \qw & \multigate{1}{\hat{U}_t(\theta_t)} & \qw & \qw & \qswap \qwx[-1] & \multigate{1}{\hat{U}_t(\theta_t/2)} & \qswap \qwx[-1] & \qw & \gate{\hat{U}_\mu(\beta_\uparrow/2)} & \qw & \multigate{1}{\hat{U}_U(\theta_U)} & \qw & \gate{\hat{U}_\mu(\beta_\uparrow/2)} & \qw & \qswap \qwx[-1] & \multigate{1}{\hat{U}_t(\theta_t/2)} & \qswap \qwx[-1] & \qw & \qw & \multigate{1}{\hat{U}_t(\theta_t)} & \qw & \qw & \qswap \qwx[-1] & \multigate{1}{\hat{U}_t(\theta_t/2)} & \qswap \qwx[-1] & \qw & \gate{\hat{U}_\mu(\beta_\uparrow/2)} & \qw & \multigate{1}{\hat{U}_U(\theta_U/2)} & \qw & \qw & \qw \\
\lstick{q_3} & \ghost{\hat{U}_U(\theta_U/2)} & \qw & \gate{\hat{U}_\mu(\beta_\downarrow/2)} & \qw & \qw & \ghost{\hat{U}_t(\theta_t/2)} & \qw & \qw & \qswap & \ghost{\hat{U}_t(\theta_t)} & \qswap & \qw & \qw & \ghost{\hat{U}_t(\theta_t/2)} & \qw & \qw & \gate{\hat{U}_\mu(\beta_\downarrow/2)} & \qw & \ghost{\hat{U}_U(\theta_U)} & \qw & \gate{\hat{U}_\mu(\beta_\downarrow/2)} & \qw & \qw & \ghost{\hat{U}_t(\theta_t/2)} & \qw & \qw & \qswap & \ghost{\hat{U}_t(\theta_t)} & \qswap & \qw & \qw & \ghost{\hat{U}_t(\theta_t/2)} & \qw & \qw & \gate{\hat{U}_\mu(\beta_\downarrow/2)} & \qw & \ghost{\hat{U}_U(\theta_U)} & \qw & \gate{\hat{U}_\mu(\beta_\downarrow/2)} & \qw & \qw & \ghost{\hat{U}_t(\theta_t/2)} & \qw & \qw & \qswap & \ghost{\hat{U}_t(\theta_t)} & \qswap & \qw & \qw & \ghost{\hat{U}_t(\theta_t/2)} & \qw & \qw & \gate{\hat{U}_\mu(\beta_\downarrow/2)} & \qw & \ghost{\hat{U}_U(\theta_U/2)} & \qw & \qw & \qw \\
\lstick{q_4} & \multigate{1}{\hat{U}_U(\theta_U/2)} & \qw & \gate{\hat{U}_\mu(\beta_\uparrow/2)} & \qw & \qw & \multigate{1}{\hat{U}_t(\theta_t/2)} & \qw & \qw & \qswap \qwx[-1] & \multigate{1}{\hat{U}_t(\theta_t)} & \qswap \qwx[-1] & \qw & \qw & \multigate{1}{\hat{U}_t(\theta_t/2)} & \qw & \qw & \gate{\hat{U}_\mu(\beta_\uparrow/2)} & \qw & \multigate{1}{\hat{U}_U(\theta_U)} & \qw & \gate{\hat{U}_\mu(\beta_\uparrow/2)} & \qw & \qw & \multigate{1}{\hat{U}_t(\theta_t/2)} & \qw & \qw & \qswap \qwx[-1] & \multigate{1}{\hat{U}_t(\theta_t)} & \qswap \qwx[-1] & \qw & \qw & \multigate{1}{\hat{U}_t(\theta_t/2)} & \qw & \qw & \gate{\hat{U}_\mu(\beta_\uparrow/2)} & \qw & \multigate{1}{\hat{U}_U(\theta_U)} & \qw & \gate{\hat{U}_\mu(\beta_\uparrow/2)} & \qw & \qw & \multigate{1}{\hat{U}_t(\theta_t/2)} & \qw & \qw & \qswap \qwx[-1] & \multigate{1}{\hat{U}_t(\theta_t)} & \qswap \qwx[-1] & \qw & \qw & \multigate{1}{\hat{U}_t(\theta_t/2)} & \qw & \qw & \gate{\hat{U}_\mu(\beta_\uparrow/2)} & \qw & \multigate{1}{\hat{U}_U(\theta_U/2)} & \qw & \qw & \qw \\
\lstick{q_5} & \ghost{\hat{U}_U(\theta_U/2)} & \qw & \gate{\hat{U}_\mu(\beta_\downarrow/2)} & \qw & \qswap & \ghost{\hat{U}_t(\theta_t/2)} & \qswap & \qw & \qw & \qw & \qw & \qw & \qswap & \ghost{\hat{U}_t(\theta_t/2)} & \qswap & \qw & \gate{\hat{U}_\mu(\beta_\downarrow/2)} & \qw & \ghost{\hat{U}_U(\theta_U)} & \qw & \gate{\hat{U}_\mu(\beta_\downarrow/2)} & \qw & \qswap & \ghost{\hat{U}_t(\theta_t/2)} & \qswap & \qw & \qw & \qw & \qw & \qw & \qswap & \ghost{\hat{U}_t(\theta_t/2)} & \qswap & \qw & \gate{\hat{U}_\mu(\beta_\downarrow/2)} & \qw & \ghost{\hat{U}_U(\theta_U)} & \qw & \gate{\hat{U}_\mu(\beta_\downarrow/2)} & \qw & \qswap & \ghost{\hat{U}_t(\theta_t/2)} & \qswap & \qw & \qw & \qw & \qw & \qw & \qswap & \ghost{\hat{U}_t(\theta_t/2)} & \qswap & \qw & \gate{\hat{U}_\mu(\beta_\downarrow/2)} & \qw & \ghost{\hat{U}_U(\theta_U/2)} & \qw & \qw & \qw \\
\lstick{q_6} & \multigate{1}{\hat{U}_U(\theta_U/2)} & \qw & \gate{\hat{U}_\mu(\beta_\uparrow/2)} & \qw & \qswap \qwx[-1] & \multigate{1}{\hat{U}_t(\theta_t/2)} & \qswap \qwx[-1] & \qw & \qw & \qw & \qw & \qw & \qswap \qwx[-1] & \multigate{1}{\hat{U}_t(\theta_t/2)} & \qswap \qwx[-1] & \qw & \gate{\hat{U}_\mu(\beta_\uparrow/2)} & \qw & \multigate{1}{\hat{U}_U(\theta_U)} & \qw & \gate{\hat{U}_\mu(\beta_\uparrow/2)} & \qw & \qswap \qwx[-1] & \multigate{1}{\hat{U}_t(\theta_t/2)} & \qswap \qwx[-1] & \qw & \qw & \qw & \qw & \qw & \qswap \qwx[-1] & \multigate{1}{\hat{U}_t(\theta_t/2)} & \qswap \qwx[-1] & \qw & \gate{\hat{U}_\mu(\beta_\uparrow/2)} & \qw & \multigate{1}{\hat{U}_U(\theta_U)} & \qw & \gate{\hat{U}_\mu(\beta_\uparrow/2)} & \qw & \qswap \qwx[-1] & \multigate{1}{\hat{U}_t(\theta_t/2)} & \qswap \qwx[-1] & \qw & \qw & \qw & \qw & \qw & \qswap \qwx[-1] & \multigate{1}{\hat{U}_t(\theta_t/2)} & \qswap \qwx[-1] & \qw & \gate{\hat{U}_\mu(\beta_\uparrow/2)} & \qw & \multigate{1}{\hat{U}_U(\theta_U/2)} & \qw & \qw & \qw \\
\lstick{q_7} & \ghost{\hat{U}_U(\theta_U/2)} & \qw & \gate{\hat{U}_\mu(\beta_\downarrow/2)} & \qw & \qw & \ghost{\hat{U}_t(\theta_t/2)} & \qw & \qw & \qw & \qw & \qw & \qw & \qw & \ghost{\hat{U}_t(\theta_t/2)} & \qw & \qw & \gate{\hat{U}_\mu(\beta_\downarrow/2)} & \qw & \ghost{\hat{U}_U(\theta_U)} & \qw & \gate{\hat{U}_\mu(\beta_\downarrow/2)} & \qw & \qw & \ghost{\hat{U}_t(\theta_t/2)} & \qw & \qw & \qw & \qw & \qw & \qw & \qw & \ghost{\hat{U}_t(\theta_t/2)} & \qw & \qw & \gate{\hat{U}_\mu(\beta_\downarrow/2)} & \qw & \ghost{\hat{U}_U(\theta_U)} & \qw & \gate{\hat{U}_\mu(\beta_\downarrow/2)} & \qw & \qw & \ghost{\hat{U}_t(\theta_t/2)} & \qw & \qw & \qw & \qw & \qw & \qw & \qw & \ghost{\hat{U}_t(\theta_t/2)} & \qw & \qw & \gate{\hat{U}_\mu(\beta_\downarrow/2)} & \qw & \ghost{\hat{U}_U(\theta_U/2)} & \qw & \qw & \qw \\
}
}
}
\caption{Quantum circuit of the optimized second-order Trotterization with three Trotter steps $r=3$.}
\label{fig:second-trotter-optimized}
\end{figure}
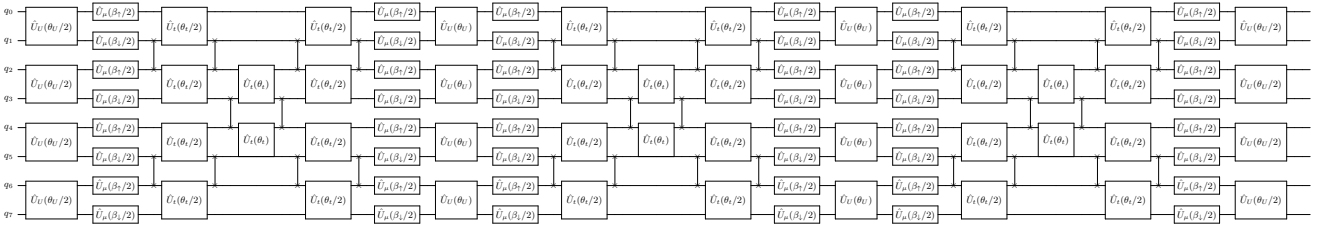

The time evolution operator can be decomposed using second-order Trotterization as,
\begin{equation}
	U^{(2)}(\tau)=\prod_{k=1}^{r}\hat{U}^{(2)}_{\mathrm{Trotter}_{k}}(\delta\tau)\,,
	\label{eq:second-trotter-1}
\end{equation}
where $r$ is the number of Trotter steps with $\delta\tau=\tau/r$ the Trotter step size. A single Trotter step in the second-order Trotterization is defined as,
\begin{eqnarray}
	\hat{U}^{(2)}_{\mathrm{Trotter}}(\delta\tau)= \hat{\mathbf{U}}_{U}\left(\frac{\theta_{U}}{2}\right).\hat{\mathbf{U}}_{\mu}\left(\frac{\beta_{\uparrow}}{2},\frac{\beta_{\downarrow}}{2}\right).\hat{\mathbf{U}}^{t}_{\mathrm{Layer-1}}\left(\frac{\theta_{t}}{2}\right).\hat{\mathbf{U}}^{t}_{\mathrm{Layer-2}}\left(\frac{\theta_{t}}{2}\right)&.&\hat{\mathbf{U}}^{t}_{\mathrm{Layer-2}}\left(\frac{\theta_{t}}{2}\right).\hat{\mathbf{U}}^{t}_{\mathrm{Layer-1}}\left(\frac{\theta_{t}}{2}\right).\nonumber\\
	&&\hat{\mathbf{U}}_{\mu}\left(\frac{\beta_{\uparrow}}{2},\frac{\beta_{\downarrow}}{2}\right).\hat{\mathbf{U}}_{U}\left(\frac{\theta_{U}}{2}\right)\,.
	\label{eq:second-trotter-2}
\end{eqnarray}
The quantum circuit associated with the single Trotter step of the second-order Trotterization is presented in Fig.~\ref{fig:single-trotter-step-2nd}.
\begin{figure}[h!]
\centerline{
\scalebox{0.85}{
\Qcircuit @C=1em @R=0.8em @!R {
    \lstick{q_0:} & \multigate{1}{\hat{U}_U(\theta_U/2)} & \qw & \gate{\hat{U}_\mu(\beta_{\uparrow}/2)} & \qw & \qw & \multigate{1}{\hat{U}_t(\theta_t/2)} & \qw & \qw & \qw & \qw & \qw & \qw & \qw & \qw & \qw & \multigate{1}{\hat{U}_t(\theta_t/2)} & \qw & \qw & \gate{\hat{U}_\mu(\beta_{\uparrow}/2)} & \qw & \multigate{1}{\hat{U}_U(\theta_U/2)} & \qw & \qw & \qw \\
    \lstick{q_1:} & \ghost{\hat{U}_U(\theta_U/2)}         & \qw & \gate{\hat{U}_\mu(\beta_{\downarrow}/2)} & \qw & \qswap & \ghost{\hat{U}_t(\theta_t/2)} & \qswap & \qw & \qw & \qw & \qw & \qw & \qw & \qw & \qswap & \ghost{\hat{U}_t(\theta_t/2)} & \qswap & \qw & \gate{\hat{U}_\mu(\beta_{\downarrow}/2)} & \qw & \ghost{\hat{U}_U(\theta_U/2)} & \qw & \qw & \qw \\
    \lstick{q_2:} & \multigate{1}{\hat{U}_U(\theta_U/2)} & \qw & \gate{\hat{U}_\mu(\beta_{\uparrow}/2)} & \qw & \qswap \qwx[-1] & \multigate{1}{\hat{U}_t(\theta_t/2)} & \qswap \qwx[-1] & \qw & \qw & \multigate{1}{\hat{U}_t(\theta_t/2)} & \qw & \multigate{1}{\hat{U}_t(\theta_t/2)} & \qw & \qw & \qswap \qwx[-1] & \multigate{1}{\hat{U}_t(\theta_t/2)} & \qswap \qwx[-1] & \qw & \gate{\hat{U}_\mu(\beta_{\uparrow}/2)} & \qw & \multigate{1}{\hat{U}_U(\theta_U/2)} & \qw & \qw & \qw \\
    \lstick{q_3:} & \ghost{\hat{U}_U(\theta_U/2)}         & \qw & \gate{\hat{U}_\mu(\beta_{\downarrow}/2)} & \qw & \qw & \ghost{\hat{U}_t(\theta_t/2)} & \qw & \qw & \qswap & \ghost{\hat{U}_t(\theta_t/2)} & \qw & \ghost{\hat{U}_t(\theta_t/2)} & \qswap & \qw & \qw & \ghost{\hat{U}_t(\theta_t/2)} & \qw & \qw & \gate{\hat{U}_\mu(\beta_{\downarrow}/2)} & \qw & \ghost{\hat{U}_U(\theta_U/2)} & \qw & \qw & \qw \\
    \lstick{q_4:} & \multigate{1}{\hat{U}_U(\theta_U/2)} & \qw & \gate{\hat{U}_\mu(\beta_{\uparrow}/2)} & \qw & \qw & \multigate{1}{\hat{U}_t(\theta_t/2)} & \qw & \qw & \qswap \qwx[-1] & \multigate{1}{\hat{U}_t(\theta_t/2)} & \qw & \multigate{1}{\hat{U}_t(\theta_t/2)} & \qswap \qwx[-1] & \qw & \qw & \multigate{1}{\hat{U}_t(\theta_t/2)} & \qw & \qw & \gate{\hat{U}_\mu(\beta_{\uparrow}/2)} & \qw & \multigate{1}{\hat{U}_U(\theta_U/2)} & \qw & \qw & \qw \\
    \lstick{q_5:} & \ghost{\hat{U}_U(\theta_U/2)}         & \qw & \gate{\hat{U}_\mu(\beta_{\downarrow}/2)} & \qw & \qswap & \ghost{\hat{U}_t(\theta_t/2)} & \qswap & \qw & \qw & \ghost{\hat{U}_t(\theta_t/2)} & \qw & \ghost{\hat{U}_t(\theta_t/2)} & \qw & \qw & \qswap & \ghost{\hat{U}_t(\theta_t/2)} & \qswap & \qw & \gate{\hat{U}_\mu(\beta_{\downarrow}/2)} & \qw & \ghost{\hat{U}_U(\theta_U/2)} & \qw & \qw & \qw \\
    \lstick{q_6:} & \multigate{1}{\hat{U}_U(\theta_U/2)} & \qw & \gate{\hat{U}_\mu(\beta_{\uparrow}/2)} & \qw & \qswap \qwx[-1] & \multigate{1}{\hat{U}_t(\theta_t/2)} & \qswap \qwx[-1] & \qw & \qw & \qw & \qw & \qw & \qw & \qw & \qswap \qwx[-1] & \multigate{1}{\hat{U}_t(\theta_t/2)} & \qswap \qwx[-1] & \qw & \gate{\hat{U}_\mu(\beta_{\uparrow}/2)} & \qw & \multigate{1}{\hat{U}_U(\theta_U/2)} & \qw & \qw & \qw \\
    \lstick{q_7:} & \ghost{\hat{U}_U(\theta_U/2)}         & \qw & \gate{\hat{U}_\mu(\beta_{\downarrow}/2)} & \qw & \qw & \ghost{\hat{U}_t(\theta_t/2)} & \qw & \qw & \qw & \qw & \qw & \qw & \qw & \qw & \qw & \ghost{\hat{U}_t(\theta_t/2)} & \qw & \qw & \gate{\hat{U}_\mu(\beta_{\downarrow}/2)} & \qw & \ghost{\hat{U}_U(\theta_U/2)} & \qw & \qw & \qw
}
}
}
\caption{Quantum circuit of the single Trotter step in the second-order Trotterization for $L=4$ fermion sites ($N=8$ qubits).}
\label{fig:single-trotter-step-2nd}
\end{figure}
 As in the case of the first-order Trotterization, the circuit depth of the quantum circuit associated with the second-order Trotterization presented in this work remains constant with increasing fermion sites $L$ or qubits $N=2L$, though the latter is significantly deeper. To reduce the circuit depth, we introduce additional optimizations to the second-order Trotterization for a given number of Trotter steps. In Fig.~\ref{fig:second-trotter-two-steps}, two Trotter steps in the second-order Trotterization for $L=4$ fermion sites are shown where the two $\hat{\mathbf{U}}^{t}_{\mathrm{Layer-2}}(\theta_{t}/2)$ quantum circuit layers in the straight-lined box within a single Trotter step can be merged to $\hat{\mathbf{U}}^{t}_{\mathrm{Layer-2}}(\theta_{t})$, and two $\hat{\mathbf{U}}_{U}(\theta_{U}/2)$ quantum circuit layers in the dotted-lined box from two consecutive Trotter steps can be merged to $\hat{\mathbf{U}}_{U}(\theta_{U})$ due to the property of unitary operator $U(\theta_{1}).U(\theta_{2})=U(\theta_{1}+\theta_{2})$.
\begin{figure}[h!]
\centerline{
\scalebox{0.50}{
\Qcircuit @C=0.7em @R=0.8em @!R {
    \lstick{q_0:} & \multigate{1}{\hat{U}_U(\theta_U/2)} & \qw & \gate{\hat{U}_\mu(\beta_{\uparrow}/2)} & \qw & \qw & \multigate{1}{\hat{U}_t(\theta_t/2)} & \qw & \qw & \qw & \qw & \qw & \qw & \qw & \qw & \qw & \multigate{1}{\hat{U}_t(\theta_t/2)} & \qw & \qw & \gate{\hat{U}_\mu(\beta_{\uparrow}/2)} & \qw & \multigate{1}{\hat{U}_U(\theta_U/2)} & \qw & \multigate{1}{\hat{U}_U(\theta_U/2)} & \qw & \gate{\hat{U}_\mu(\beta_{\uparrow}/2)} & \qw & \qw & \multigate{1}{\hat{U}_t(\theta_t/2)} & \qw & \qw & \qw & \qw & \qw & \qw & \qw & \qw & \qw & \multigate{1}{\hat{U}_t(\theta_t/2)} & \qw & \qw & \gate{\hat{U}_\mu(\beta_{\uparrow}/2)} & \qw & \multigate{1}{\hat{U}_U(\theta_U/2)} & \qw & \qw & \qw \\
    \lstick{q_1:} & \ghost{\hat{U}_U(\theta_U/2)} & \qw & \gate{\hat{U}_\mu(\beta_{\downarrow}/2)} & \qw & \qswap & \ghost{\hat{U}_t(\theta_t/2)} & \qswap & \qw & \qw & \qw & \qw & \qw & \qw & \qw & \qswap & \ghost{\hat{U}_t(\theta_t/2)} & \qswap & \qw & \gate{\hat{U}_\mu(\beta_{\downarrow}/2)} & \qw & \ghost{\hat{U}_U(\theta_U/2)} & \qw & \ghost{\hat{U}_U(\theta_U/2)} & \qw & \gate{\hat{U}_\mu(\beta_{\downarrow}/2)} & \qw & \qswap & \ghost{\hat{U}_t(\theta_t/2)} & \qswap & \qw & \qw & \qw & \qw & \qw & \qw & \qw & \qswap & \ghost{\hat{U}_t(\theta_t/2)} & \qswap & \qw & \gate{\hat{U}_\mu(\beta_{\downarrow}/2)} & \qw & \ghost{\hat{U}_U(\theta_U/2)} & \qw & \qw & \qw \\
    \lstick{q_2:} & \multigate{1}{\hat{U}_U(\theta_U/2)} & \qw & \gate{\hat{U}_\mu(\beta_{\uparrow}/2)} & \qw & \qswap \qwx[-1] & \multigate{1}{\hat{U}_t(\theta_t/2)} & \qswap \qwx[-1] & \qw & \qw & \multigate{1}{\hat{U}_t(\theta_t/2)} & \qw & \multigate{1}{\hat{U}_t(\theta_t/2)} & \qw & \qw & \qswap \qwx[-1] & \multigate{1}{\hat{U}_t(\theta_t/2)} & \qswap \qwx[-1] & \qw & \gate{\hat{U}_\mu(\beta_{\uparrow}/2)} & \qw & \multigate{1}{\hat{U}_U(\theta_U/2)} & \qw & \multigate{1}{\hat{U}_U(\theta_U/2)} & \qw & \gate{\hat{U}_\mu(\beta_{\uparrow}/2)} & \qw & \qswap \qwx[-1] & \multigate{1}{\hat{U}_t(\theta_t/2)} & \qswap \qwx[-1] & \qw & \qw & \multigate{1}{\hat{U}_t(\theta_t/2)} & \qw & \multigate{1}{\hat{U}_t(\theta_t/2)} & \qw & \qw & \qswap \qwx[-1] & \multigate{1}{\hat{U}_t(\theta_t/2)} & \qswap \qwx[-1] & \qw & \gate{\hat{U}_\mu(\beta_{\uparrow}/2)} & \qw & \multigate{1}{\hat{U}_U(\theta_U/2)} & \qw & \qw & \qw \\
    \lstick{q_3:} & \ghost{\hat{U}_U(\theta_U/2)} & \qw & \gate{\hat{U}_\mu(\beta_{\downarrow}/2)} & \qw & \qw & \ghost{\hat{U}_t(\theta_t/2)} & \qw & \qw & \qswap & \ghost{\hat{U}_t(\theta_t/2)} & \qw & \ghost{\hat{U}_t(\theta_t/2)} & \qswap & \qw & \qw & \ghost{\hat{U}_t(\theta_t/2)} & \qw & \qw & \gate{\hat{U}_\mu(\beta_{\downarrow}/2)} & \qw & \ghost{\hat{U}_U(\theta_U/2)} & \qw & \ghost{\hat{U}_U(\theta_U/2)} & \qw & \gate{\hat{U}_\mu(\beta_{\downarrow}/2)} & \qw & \qw & \ghost{\hat{U}_t(\theta_t/2)} & \qw & \qw & \qswap & \ghost{\hat{U}_t(\theta_t/2)} & \qw & \ghost{\hat{U}_t(\theta_t/2)} & \qswap & \qw & \qw & \ghost{\hat{U}_t(\theta_t/2)} & \qw & \qw & \gate{\hat{U}_\mu(\beta_{\downarrow}/2)} & \qw & \ghost{\hat{U}_U(\theta_U/2)} & \qw & \qw & \qw \\
    \lstick{q_4:} & \multigate{1}{\hat{U}_U(\theta_U/2)} & \qw & \gate{\hat{U}_\mu(\beta_{\uparrow}/2)} & \qw & \qw & \multigate{1}{\hat{U}_t(\theta_t/2)} & \qw & \qw & \qswap \qwx[-1] & \multigate{1}{\hat{U}_t(\theta_t/2)} & \qw & \multigate{1}{\hat{U}_t(\theta_t/2)} & \qswap \qwx[-1] & \qw & \qw & \multigate{1}{\hat{U}_t(\theta_t/2)} & \qw & \qw & \gate{\hat{U}_\mu(\beta_{\uparrow}/2)} & \qw & \multigate{1}{\hat{U}_U(\theta_U/2)} & \qw & \multigate{1}{\hat{U}_U(\theta_U/2)} & \qw & \gate{\hat{U}_\mu(\beta_{\uparrow}/2)} & \qw & \qw & \multigate{1}{\hat{U}_t(\theta_t/2)} & \qw & \qw & \qswap \qwx[-1] & \multigate{1}{\hat{U}_t(\theta_t/2)} & \qw & \multigate{1}{\hat{U}_t(\theta_t/2)} & \qswap \qwx[-1] & \qw & \qw & \multigate{1}{\hat{U}_t(\theta_t/2)} & \qw & \qw & \gate{\hat{U}_\mu(\beta_{\uparrow}/2)} & \qw & \multigate{1}{\hat{U}_U(\theta_U/2)} & \qw & \qw & \qw \\
    \lstick{q_5:} & \ghost{\hat{U}_U(\theta_U/2)} & \qw & \gate{\hat{U}_\mu(\beta_{\downarrow}/2)} & \qw & \qswap & \ghost{\hat{U}_t(\theta_t/2)} & \qswap & \qw & \qw & \ghost{\hat{U}_t(\theta_t/2)} & \qw & \ghost{\hat{U}_t(\theta_t/2)} & \qw & \qw & \qswap & \ghost{\hat{U}_t(\theta_t/2)} & \qswap & \qw & \gate{\hat{U}_\mu(\beta_{\downarrow}/2)} & \qw & \ghost{\hat{U}_U(\theta_U/2)} & \qw & \ghost{\hat{U}_U(\theta_U/2)} & \qw & \gate{\hat{U}_\mu(\beta_{\downarrow}/2)} & \qw & \qswap & \ghost{\hat{U}_t(\theta_t/2)} & \qswap & \qw & \qw & \ghost{\hat{U}_t(\theta_t/2)} & \qw & \ghost{\hat{U}_t(\theta_t/2)} & \qw & \qw & \qswap & \ghost{\hat{U}_t(\theta_t/2)} & \qswap & \qw & \gate{\hat{U}_\mu(\beta_{\downarrow}/2)} & \qw & \ghost{\hat{U}_U(\theta_U/2)} & \qw & \qw & \qw \\
    \lstick{q_6:} & \multigate{1}{\hat{U}_U(\theta_U/2)} & \qw & \gate{\hat{U}_\mu(\beta_{\uparrow}/2)} & \qw & \qswap \qwx[-1] & \multigate{1}{\hat{U}_t(\theta_t/2)} & \qswap \qwx[-1] & \qw & \qw & \qw & \qw & \qw & \qw & \qw & \qswap \qwx[-1] & \multigate{1}{\hat{U}_t(\theta_t/2)} & \qswap \qwx[-1] & \qw & \gate{\hat{U}_\mu(\beta_{\uparrow}/2)} & \qw & \multigate{1}{\hat{U}_U(\theta_U/2)} & \qw & \multigate{1}{\hat{U}_U(\theta_U/2)} & \qw & \gate{\hat{U}_\mu(\beta_{\uparrow}/2)} & \qw & \qswap \qwx[-1] & \multigate{1}{\hat{U}_t(\theta_t/2)} & \qswap \qwx[-1] & \qw & \qw & \qw & \qw & \qw & \qw & \qw & \qswap \qwx[-1] & \multigate{1}{\hat{U}_t(\theta_t/2)} & \qswap \qwx[-1] & \qw & \gate{\hat{U}_\mu(\beta_{\uparrow}/2)} & \qw & \multigate{1}{\hat{U}_U(\theta_U/2)} & \qw & \qw & \qw \\
    \lstick{q_7:} & \ghost{\hat{U}_U(\theta_U/2)} & \qw & \gate{\hat{U}_\mu(\beta_{\downarrow}/2)} & \qw & \qw & \ghost{\hat{U}_t(\theta_t/2)} & \qw & \qw & \qw & \qw & \qw & \qw & \qw & \qw & \qw & \ghost{\hat{U}_t(\theta_t/2)} & \qw & \qw & \gate{\hat{U}_\mu(\beta_{\downarrow}/2)} & \qw & \ghost{\hat{U}_U(\theta_U/2)} & \qw & \ghost{\hat{U}_U(\theta_U/2)} & \qw & \gate{\hat{U}_\mu(\beta_{\downarrow}/2)} & \qw & \qw & \ghost{\hat{U}_t(\theta_t/2)} & \qw & \qw & \qw & \qw & \qw & \qw & \qw & \qw & \qw & \ghost{\hat{U}_t(\theta_t/2)} & \qw & \qw & \gate{\hat{U}_\mu(\beta_{\downarrow}/2)} & \qw & \ghost{\hat{U}_U(\theta_U/2)} & \qw & \qw & \qw
\gategroup{1}{10}{8}{14}{.8em}{-}
\gategroup{1}{22}{8}{24}{.8em}{--}
\gategroup{1}{32}{8}{36}{.8em}{-}
}
}
}
\caption{Quantum circuit of the second-order Trotterization with two Trotter steps $r=2$ for $L=4$ fermion sites ($N=8$ qubits).}
\label{fig:second-trotter-two-steps}
\end{figure}
The merging of such quantum circuit layers leads to an optimized second-order Trotterization,
\begin{equation}
	U^{(2)}_{\mathrm{Trotter}}=\hat{\mathrm{U}}_{U}\left(\frac{\theta_{U}}{2}\right).\left(\prod_{k=1}^{r-1}\tilde{\mathbf{U}}(\theta_{t},\beta_{\uparrow},\beta_{\downarrow}).\hat{\mathrm{U}}_{U}(\theta_{U})\right).\tilde{\mathbf{U}}(\theta_{t},\beta_{\uparrow},\beta_{\downarrow}).\hat{\mathrm{U}}_{U}\left(\frac{\theta_{U}}{2}\right)\,,
	\label{eq:second-trotter-opt}
\end{equation}
where $\tilde{\mathbf{U}}(\theta_{t},\beta_{\uparrow},\beta_{\downarrow})$ is defined as
\begin{equation}
	\tilde{\mathbf{U}}(\theta_{t},\beta_{\uparrow},\beta_{\downarrow})=\hat{\mathbf{U}}_{\mu}\left(\frac{\beta_{\uparrow}}{2},\frac{\beta_{\downarrow}}{2}\right).\hat{\mathbf{U}}^{t}_{\mathrm{Layer-1}}\left(\frac{\theta_{t}}{2}\right).\hat{\mathbf{U}}^{t}_{\mathrm{Layer-2}}(\theta_{t}).\hat{\mathbf{U}}^{t}_{\mathrm{Layer-1}}\left(\frac{\theta_{t}}{2}\right).\hat{\mathbf{U}}_{\mu}\left(\frac{\beta_{\uparrow}}{2},\frac{\beta_{\downarrow}}{2}\right)\,,
	\label{eq:second-trotter-sub}
\end{equation}
and shown in Fig.~\ref{fig:second-trotter-optimized} for $r=3$ Trotter steps.

\begin{figure}[h!]
	\centerline{\includegraphics[width=10cm]{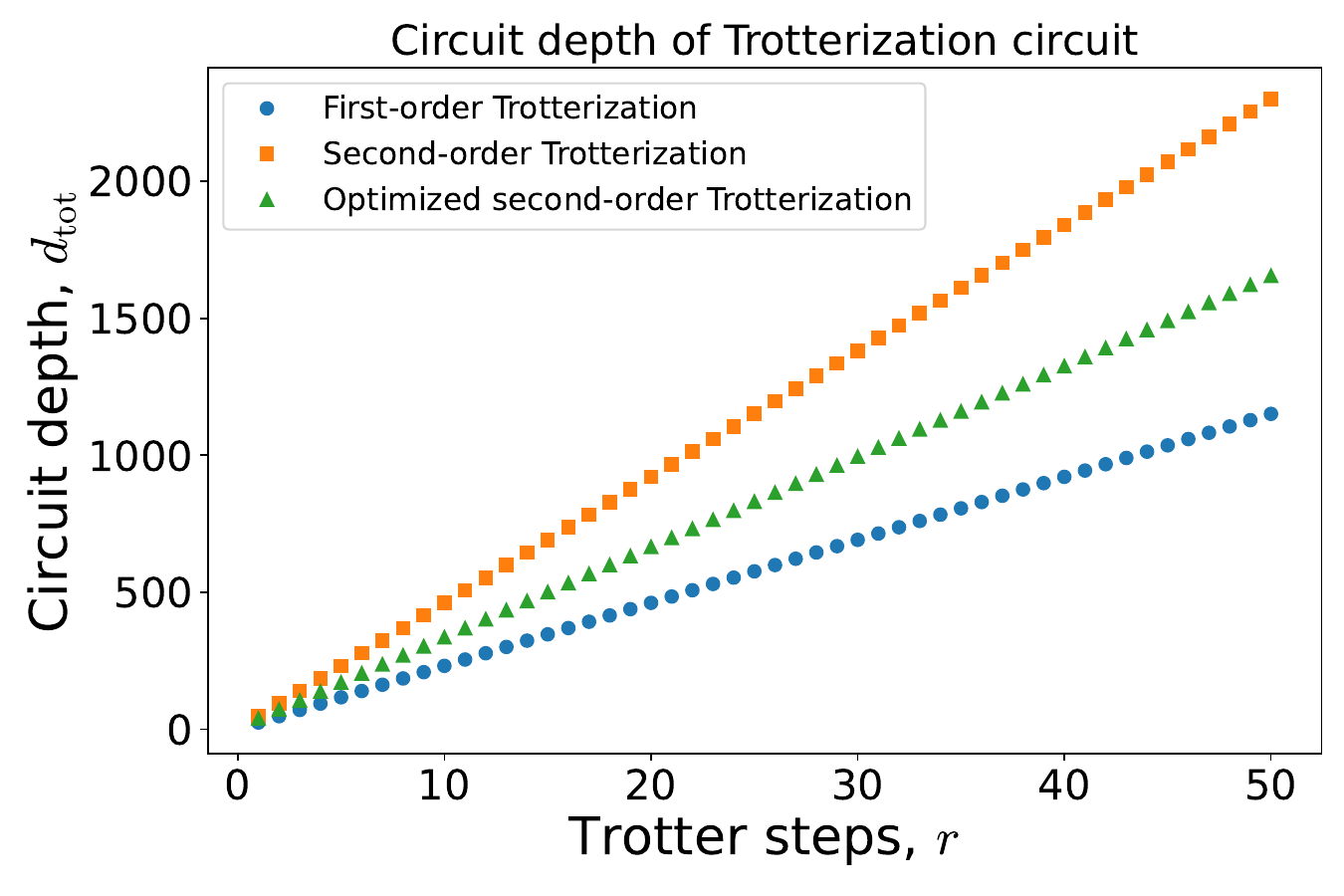}}
	\caption{The total circuit depth with respect to the number of Trotter step for the quantum circuit of the first-order, second-order, and optimized second-order Trotterization describing the time evolution driven by the Fermi-Hubbard model.}
	\label{fig:circuit-depth}
\end{figure}

In addition, the circuit depth is an important measure of how many operations one can implement in parallel on the qubit register of a quantum computer. As contemporary quantum computers have finite coherence time, the circuit depth will essentially capture the reliability of our Trotterization circuit being run on them. Based on our quantum circuit implementations and two-qubit quantum gates $\hat{U}_{t}(\theta_{t})$, $\hat{U}_{U}(\theta_{U})$ and $\hat{U}_{\mu}(\beta_{\uparrow},\beta_{\downarrow})$, the circuit depths associated with the quantum circuit of the single Trotter step for the first-order Trotterization (Fig.~\ref{fig:single-trotter-step}) and second-order Trotterization (Fig.~\ref{fig:single-trotter-step-2nd}) are $d^{(1)}_{\mathrm{single}}=23$ and $d^{(2)}_{\mathrm{single}}=46$, respectively, and remain constant as a function of the fermion site number $L$ or qubit numbers $N=2L$. Here, we only consider the quantum operations presented in the Trotterization circuits. Therefore, the total circuit depth for $r$ Trotter steps scales for first-order and second-order Trotterization circuits as $d^{(1)}_{\mathrm{tot}}=rd^{(1)}_{\mathrm{single}}$ and $d^{(2)}_{\mathrm{tot}}=rd^{(2)}_{\mathrm{single}}$, respectively. Consequently, the total circuit depth becomes substantially larger for the second-order Trotterization compared to the first-order as it scales as $d^{(2)}_{\mathrm{tot}}=2d^{(1)}_{\mathrm{tot}}$. On the other hand, the optimized second-order Trotterization circuit that we introduce in Eq.~(\ref{eq:second-trotter-opt}) provides a smaller total circuit depth for $r$ number of Trotter steps, as shown in Fig.~\ref{fig:circuit-depth}.

So far, we have developed Trotterization circuits for the Fermi-Hubbard model with open boundary conditions. However, when considering periodic boundary conditions, the hopping term between the first and $L$-th lattice sites under JW transformation leads to additional terms such as $XZZ..ZX + YZZ..ZY$ in the JW-transformed Fermi-Hubbard Hamiltonian Eq.~(\ref{eq:FHM-qubit}). These terms involve $Z$-strings acting on intermediate qubits, resulting in a circuit depth of $O(N)$ for each Trotter step. In contrast, the Trotterization circuits we developed for the Fermi-Hubbard model with open boundary conditions do not depend on the system size or total number of qubits, leading to a total circuit depth that is independent of these factors. However, for periodic boundary conditions, the total circuit depth now has an additional polynomial dependence on the number of qubits. As a result, for larger systems, this increase in total circuit depth for Trotterization significantly limits the simulation time, particularly since real quantum computers have finite coherence times. Consequently, we have chosen not to simulate the Fermi-Hubbard model with periodic boundary conditions in this work, reserving it for a future study.

\section{Experimental Realizations on IBM Quantum Computers}\label{sec:implementation}
We study the real-time dynamics of the one-dimensional Fermi-Hubbard model with open boundary conditions for $L=10$ sites ($N=20$ qubits) and $L=52$ sites ($N=104$ qubits) using IBM quantum computers. The couplings related to hopping and the on-site interaction are set to $t=1$ and $U=1$, respectively. Additionally, for simplicity, we work with the model that assumes a zero chemical potential, $\mu_{j, a}=0$ for all sites $j$ and spin states $a$. Our observable is the N\'eel observable defined in Eq.(~\ref{eq:Neel-observable}), which captures the relaxation of the staggered magnetization in the Fermi-Hubbard chain with time. For both cases with $L=10$ sites ($N=20$ qubits) and $L=52$ sites ($N=104$ qubits), we evolve the initial N\'eel state, $|\psi_{\text{N\'eel}}\rangle$, for time $\tau=5$, given in the units of $\hbar/E$, where $E$ has the dimension of energy and associated with intrinsic energy scale of the Fermi-Hubbard Hamiltonian. We use $r=10$ Trotter steps with Trotter step-size being $\delta\tau=0.5$ in both first-order and the second-order Trotterization simulations. We chose the N\'eel state as the initial state to minimize the circuit depth, as it requires only one circuit depth and is not an eigenstate of the Fermi-Hubbard Hamiltonian. Such a choice enables us to avoid the larger circuit depth associated with any other initial-state preparation circuit, allowing us to carry out as many number of Trotter steps as possible and limiting the effects of device noise.

\begin{figure}[h!]
    \centering
    \begin{subfigure}[b]{0.3\textwidth}
        \centering
        \includegraphics[width=\textwidth]{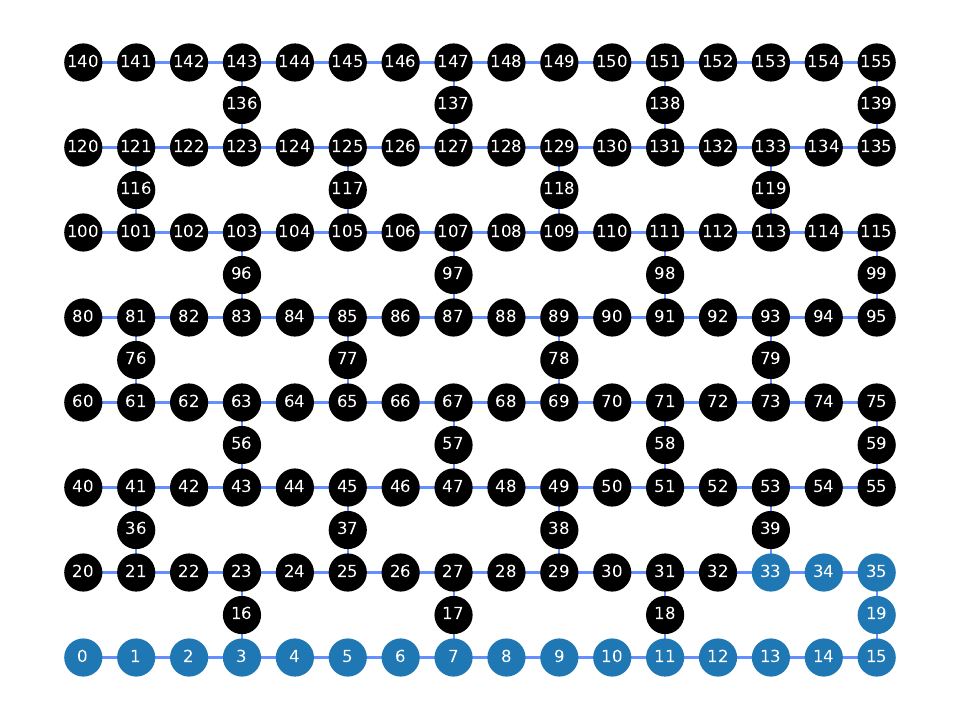}
        \caption{\texttt{ibm\_kingston}, $L =10$ sites ($N=20$ qubits)}
        \label{fig:marrakesh_10}
    \end{subfigure}
    \hfill
    \begin{subfigure}[b]{0.3\textwidth}
        \centering
        \includegraphics[width=\textwidth]{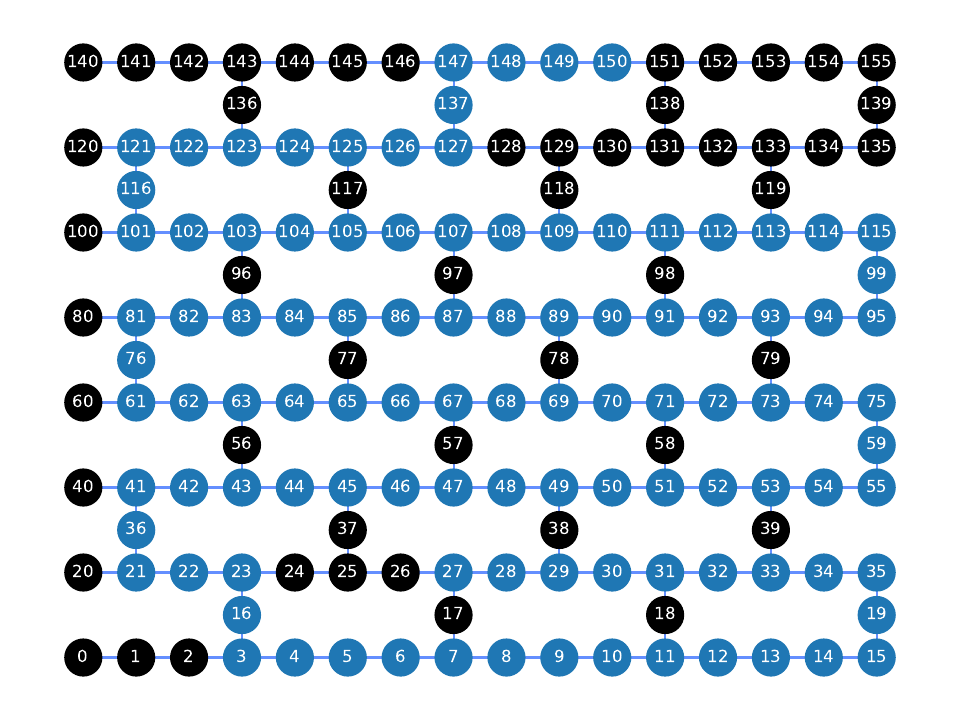}
        \caption{\texttt{ibm\_marrakesh}, $L=52$ sites ($N=104$ qubits)}
        \label{fig:marrakesh_52}
    \end{subfigure}
    \hfill
    \begin{subfigure}[b]{0.3\textwidth}
        \centering
        \includegraphics[width=\textwidth]{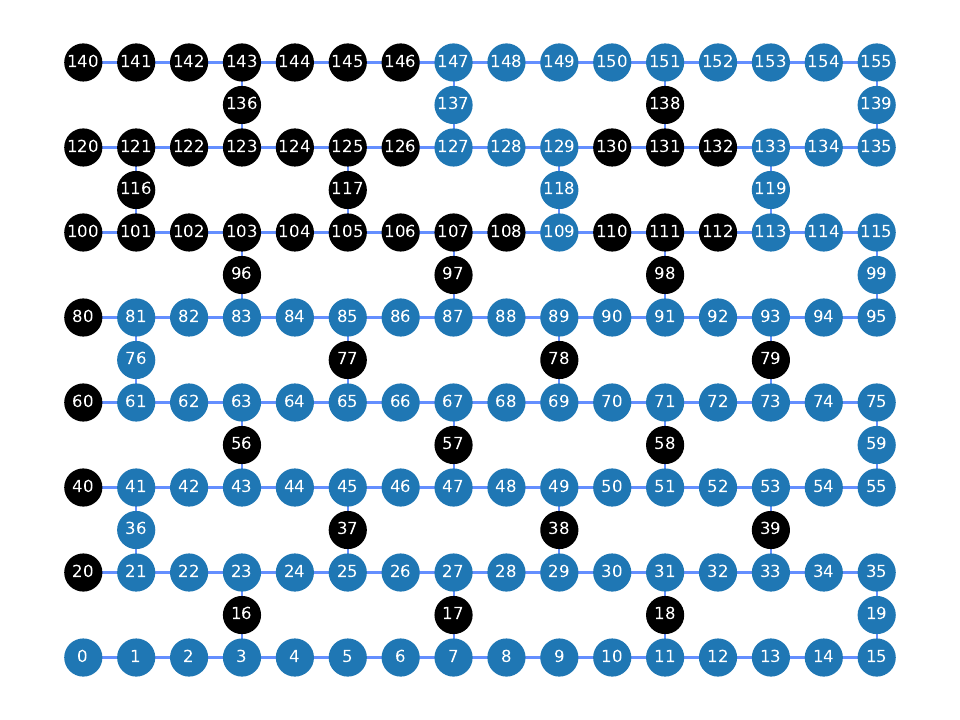}
        \caption{\texttt{ibm\_kingston}, $L=52$ sites ($N=104$ qubits)}
        \label{fig:kingston_52}
    \end{subfigure}
    \caption{The qubit mapping representing the 1D Fermi-Hubbard chain with open boundary conditions on \texttt{ibm$\_$kingston} and \texttt{ibm$\_$marrakesh}. Active (used) qubits are represented in blue, while inactive (unused) qubits are shown in black. The figures were generated using the \texttt{plot$\_$gate$\_$map()} method from \texttt{qiskit$\_$device$\_$benchmarking}~\cite{qiskit-benchmarking}. }
    \label{fig:qubit-mapping-marrakesh}
\end{figure}

The experiments are run on IBM Quantum's \texttt{ibm$\_$marrakesh} and \texttt{ibm$\_$kingston}, Heron r2 processors with 156 fixed-frequency transmon qubits and tunable couplers and heavy-hexagonal lattice topologies. Both quantum processing units (QPUs) have the basis gate set consisting of single-qubit gates
\texttt{id},
\texttt{rx},
\texttt{rz},
\texttt{x} and
\texttt{sx}; and the two-qubit gates
\texttt{cz} and
\texttt{rzz}. Properties associated to these two QPUs acquired during experimentation are given in Table~\ref{tab:QPU-parameters}. Figure~\ref{fig:qubit-mapping-marrakesh} presents the qubit mappings of our chosen one-dimensional Fermi-Hubbard systems on the \texttt{ibm$\_$marrakesh} and \texttt{ibm$\_$kingston} for different numbers of sites, $L$.

\begin{center}
\begin{table}
\begin{tabular}{lcc}
\toprule
\textbf{Parameters} & \texttt{ibm$\_$marrakesh} & \texttt{ibm$\_$kingston} \\
\midrule
Median energy relaxation time, $T_1$ & 195.74 $\mu$s & 272.99 $\mu$s\\
Median dephasing time, $T_2$ & 106.81 $\mu$s & 135.96 $\mu$s\\
Median \texttt{sx} error rate (single-qubit gate) & $2.834\times 10^{-4}$ & $2.515\times 10^{-4}$ \\
Median \texttt{cz} error rate (two-qubit gate) & $2.521\times 10^{-3}$ & $1.931\times 10^{-3}$ \\
Median readout error rate & $8.972\times 10^{-3}$ & $8.667\times 10^{-3}$ \\
\bottomrule
\end{tabular}
\caption{The parameters for \texttt{ibm$\_$marrakesh} and \texttt{ibm$\_$kingston}.}
\label{tab:QPU-parameters}
\end{table}
\end{center}

\subsection{Circuit depth and two-qubit gate count}
Table~\ref{tab:FHM-statistics-1st} and Table~\ref{tab:FHM-statistics-2nd} present the total circuit depth, the \texttt{cz} depth that ignores the single-qubit operations executed in parallel on the qubit register, and the number of \texttt{cz} gates associated after each value of total Trotter steps in our first-order and optimized second-order Trotterization circuit implementation on the \texttt{ibm$\_$marrakesh} and \texttt{ibm$\_$kingston}, after transpiling the Trotterization circuits on their basis gate sets. We see that the circuit depth and \texttt{cz} depth are close to each other for $L=10$ sites ($N=20$ qubits) and $L=52$ sites ($N=104$ qubits) after $r$ Trotter steps, which reaffirms the scalability of our Trotterization circuit mentioned in Section~\ref {sec:trotterization}.

\begin{table}[h!]
    \centering
    \caption{First-order Trotterization}
    \begin{tabular}{c|c|c|c|c|c|c}
        \hline
        & \multicolumn{3}{c|}{$L=10$ sites ($N = 20$ qubits) at \texttt{ibm$\_$kingston}} & \multicolumn{3}{c}{$L=52$ sites ($N = 104$ qubits) at \texttt{ibm$\_$marrakesh}} \\
        \cline{2-7}
        Trotter Step number & Circuit depth & \texttt{cz} depth & No. of \texttt{cz} & Circuit depth & \texttt{cz} depth & No. of \texttt{cz} \\
        \hline
        \hline
        1 & 50 & 19 & 98 & 53 & 19 & 539 \\
        \hline
        2 & 102 & 37 & 208 & 99 & 34 & 1153 \\
        \hline
        3 & 150 & 52 & 318 & 157 & 55 & 1767 \\
        \hline
        4 & 208 & 73 & 428 & 209 & 73 & 2381 \\
        \hline
        5 & 260 & 91 & 538 & 255 & 88 & 2995 \\
        \hline
        6 & 306 & 106 & 648 & 313 & 109 & 3609 \\
        \hline
        7 & 362 & 127 & 758 & 359 & 124 & 4223 \\
        \hline
        8 & 416 & 145 & 868 & 417 & 145 & 4837 \\
        \hline
        9 & 466 & 163 & 978 & 469 & 163 & 5451 \\
        \hline
        10 & 518 & 181 & 1088 & 515 & 178 & 6065 \\
        \hline
    \end{tabular}
    \vspace{0.2cm}
    \begin{flushleft}
    The circuit depth, \texttt{cz} depth, and number of \texttt{cz} gates with number of Trotter steps in first-order Trotterization for $L=10$ sites ($N=20$ qubits) and $L=52$ sites ($N=104$ qubits) at \texttt{ibm$\_$kingston} and \texttt{ibm$\_$marrakesh}, respectively, after transpilation with optimization level 3 in Qiskit.
    \end{flushleft}
    \label{tab:FHM-statistics-1st}
\end{table}

\begin{table}[h!]
    \centering
    \caption{Optimized Second-order Trotterization}
    \begin{tabular}{c|c|c|c|c|c|c}
        \hline
        & \multicolumn{3}{c|}{$L=10$ sites ($N = 20$ qubits)} & \multicolumn{3}{c}{$L=52$ sites ($N = 104$ qubits)} \\
        \cline{2-7}
        Trotter Step number & Circuit depth & \texttt{cz} depth & No. of \texttt{cz} & Circuit depth & \texttt{cz} depth & No. of \texttt{cz} \\
        \hline
        \hline
        1 & 93 & 29 & 180 & 95 & 29 & 978 \\
        \hline
        2 & 176 & 55 & 340 & 180 & 55 & 1852 \\
        \hline
        3 & 259 & 81 & 500 & 265 & 81 & 2726 \\
        \hline
        4 & 342 & 107 & 660 & 350 & 107 & 3600 \\
        \hline
        5 & 425 & 133 & 820 & 435 & 133 & 4474 \\
        \hline
        6 & 508 & 159 & 980 & 520 & 159 & 5348 \\
        \hline
        7 & 591 & 185 & 1140 & 605 & 185 & 6222 \\
        \hline
        8 & 674 & 211 & 1300 & 690 & 211 & 7096 \\
        \hline
        9 & 757 & 237 & 1460 & 775 & 237 & 7970 \\
        \hline
        10 & 840 & 263 & 1620 & 860 & 263 & 8844 \\
        \hline
    \end{tabular}
    \vspace{0.2cm}
    \begin{flushleft}
    The circuit depth, \texttt{cz} depth, and number of \texttt{cz} gates with number of Trotter steps in optimized second-order Trotterization for both $L=10$ sites ($N=20$ qubits) and $L=52$ sites ($N=104$ qubits), at \texttt{ibm$\_$kingston}, after transpilation with optimization level 3 in Qiskit.
    \end{flushleft}
    \label{tab:FHM-statistics-2nd}
\end{table}

\section{Result and Discussion}\label{sec:discussion}

We benchmark the accuracy of our time evolution experiments on the IBM quantum computers using a classical exact method for the small system size ($L=10$ sites or $N=20$ qubits) and a classical approximation method based on Matrix product state (MPS) and time-dependent variational principle (TDVP) for the large-scale system ($L=52$ sites or $N=104$ qubits). Besides, we set $t=1$ and $U=1$ in this work. Choosing large $U\gg |t|$ would lead to the $t-J$ model where there is no double occupancy at a single site~\cite{Essler-Frahm-Gohmann-Klumper-Korepin-2005}. Furthermore, for strong repulsion, $U\gg |t|$ and at half-filling where the total number of electrons $N_{\mathrm{tot}}$ equals the number of lattice sites $L$ and a single lattice site is occupied precisely by a single electron, the excitations of an antiferromagnetic spin chain effectively describe the low-lying excitations of the Fermi-Hubbard model. Also, we limit ourselves to the repulsive regime of the Fermi-Hubbard model, $U>0$, to avoid the quantum phase transition at $U=0$.

The exact method is the straightforward computation of the time-evolved expectation value of the N\'eel observable $\langle\psi(\tau)|\hat{O}_{\text{N\'eel}}|\psi(\tau)\rangle$ using the vector-matrix-vector product of conjugate state vector, operator, and the state vector. The evolved quantum state is $|\psi(\tau)\rangle = e^{-i H \tau/\hbar}|\psi_{\text{N\'eel}}\rangle$, which is determined by the matrix exponentiation of the Hamiltonian $H$, a $4^{L}\times 4^{L}$ Hermitian matrix acting on the Hilbert space of dimension $4^{L}$. We have implemented
this approach using QuSpin~\cite{quspin}. This method is the simplest and most accurate, however, becomes computationally intractable for calculating the time evolution of state vectors with $N>40$ qubits due to the Hilbert space’s dimensionality increasing exponentially with the number of qubits.
Therefore, we validate our qubit mapping, Trotterization circuits, and real device computations using the exact method only for $L=10$ sites.

\subsection{$L=10$ sites (20 qubits)}\label{sec:L10sites}

\begin{figure}[h!]
	\centerline{\includegraphics[width=10cm]{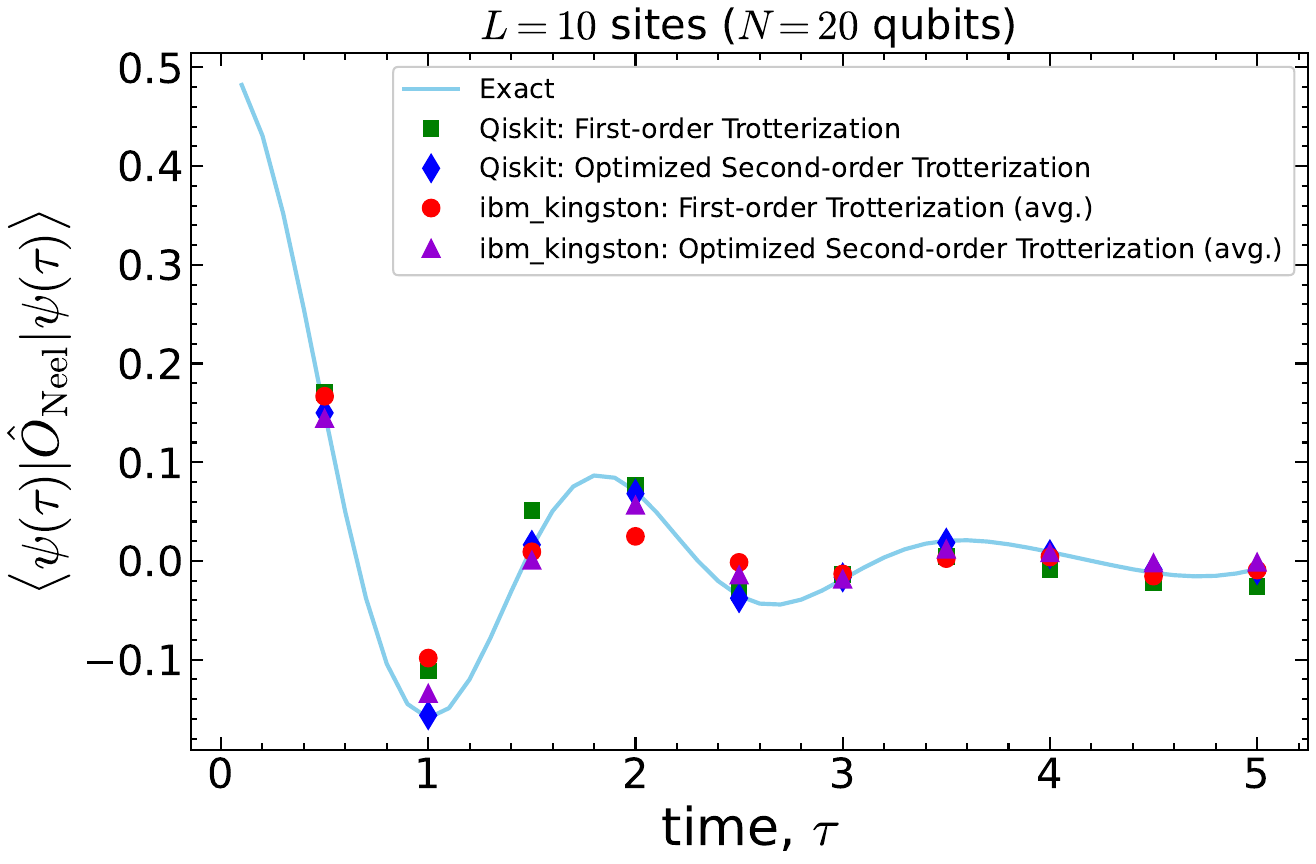}}
	\caption{The time evolution of the N\'eel observable $\langle\psi(\tau)|\hat{O}_{\text{N\'eel}}|\psi(\tau)\rangle$ for $L=10$ sites ($N=20$ qubits) with open boundary conditions where $|\psi(\tau)\rangle = e^{-i H\tau/\hbar}|\psi_{\text{N\'eel}}\rangle$. Here, the simulation time $\tau$ is in the unit $\hbar/E$ where $E$ is the intrinsic energy scale of the Fermi-Hubbard Hamiltonian. The Trotter step-size is $\delta\tau=0.5$. Here, we report the averaged value of the N\'eel observable using five instances at each value of total Trotter steps.}
	\label{fig:Neel-observable-10}
\end{figure}

In Fig.~\ref{fig:Neel-observable-10}, we can see that the expectation values of the N\'eel observable computed with respect to the time-evolved state $|\psi(\tau)\rangle$ from the initial N\'eel state $|\psi_{\text{N\'eel}}\rangle$ using \texttt{ibm$\_$kingston} agree well with the exact method and the noiseless Qiskit simulation for the Fermi-Hubbard model with $L=10$ sites. Here, we implement both the first-order Trotterization and optimized second-order Trotterization for the time evolution and the quantum error mitigation (QEM) methods detailed in App.~\ref {app:error_mitigations} to mitigate the error due to the noisy quantum computer. 
In addition, we report the averaged value of the expectation value of the N\'eel observable after five runs at each value of total Trotter steps, as can be seen from Table~\ref{tab:neel-observable}. Besides, repeating the experiment, particularly by implementing the Trotterization circuits alongside quantum error mitigation techniques, to accurately determine the expectation value for different total numbers of Trotter steps, demonstrates the correctness and reliability of our approach, as indicated by the low standard deviation observed in both the first-order and optimized second-order Trotterization circuits in Table~\ref{tab:neel-observable}. Additionally, the results from the second-order Trotterization demonstrate greater accuracy than those from the first-order Trotterization, as shown in Fig.~\ref{fig:Neel-observable-10}.

\begin{table}[ht]
    \centering
    \begin{tabular}{ccc}
        \toprule
        Trotter step number & \multicolumn{2}{c}{\textbf{Expectation value of the N\'eel observable}} \\
        \cmidrule(lr){2-3}
        & First-order Trotterization & Optimized second-order Trotterization \\
        \midrule
        1 & $0.16683985\pm 0.00529411$ & $0.14742564\pm 0.00601736$ \\
        2 & $-0.09839842\pm 0.00179805$ & $-0.1236192\pm 0.00493785$ \\
        3 & $0.00926564\pm 0.00197287$ & $-0.00574502\pm 0.00310491$ \\
        4 & $0.02491771\pm 0.00153204$ & $0.05110036\pm 0.00294584$ \\
        5 & $-0.0014325\pm 0.00128873$ & $-0.00982269\pm 0.00376725$ \\
        6 & $-0.01319128\pm 0.00217597$ & $-0.01471911\pm 0.00110681$ \\
        7 & $0.0023625\pm 0.00113029$ & $0.00956843\pm 0.00094424$ \\
        8 & $0.00397195\pm 0.00067883$ & $0.00488513\pm 0.00115636$ \\
        9 & $-0.01536343\pm 0.00128149$ & $-0.00222088\pm 0.00065659$ \\
        10 & $0.00926116\pm 0.00084341$ & $0.0010144 \pm 0.00072793$ \\
        \bottomrule
    \end{tabular}
    \caption{Average expectation value (with $\pm$ terms representing the standard deviation) of the N\'eel observable for $L=10$ sites ($N=20$ qubits) after five instances at each value of the total Trotter steps at \texttt{ibm$\_$kingston}.}
    \label{tab:neel-observable}
\end{table}

\subsection{$L=52$ sites (104 qubits)}\label{sec:L52sites}
The simulation results for the Fermi-Hubbard model with $L=52$ sites from the IBM quantum computers are compared against classical time evolution methods based on the matrix product states (MPS)~\cite{Paeckel:2019yjf, Cirac:2020obd}. Unlike classical approximation methods such as the Quantum Monte Carlo (QMC) method, which face significant limitations in time evolution computations due to the onset of the dynamical sign problem, as pointed out in Ref.~\cite{Ortiz-Gubernatis-Knill-Laflamme}, the Matrix Product State (MPS)-based method does not suffer from these constraints. Therefore, we have chosen MPS-based time evolution computation as our classical verifier for results from large-scale real quantum computers, a task that exceeds the capabilities of classical exact methods.

The MPS is a one-dimensional array of tensors linked together, with each tensor corresponding to a site or particle of the many-body system. The indices connecting the tensors in the MPS are called bond indices, which can take up to $\chi$ values (also known as bond dimensions). The open indices of each tensor correspond to the physical degrees of freedom of the local Hilbert space associated with a site or a
particle of the system, which can take up to $d$ values (for our fermionic system, $d = 4$). While the MPS can represent any quantum state of the many-body system, the bond dimension
$\chi$ needs to be exponentially large in the system size to cover all states in the Hilbert space. The time evolution of the MPS is determined using the time-dependent variational principle (TDVP)~\cite{TDVP-ref-1, TDVP-ref-2}, which we dubbed the \textit{MPS-TDVP} method, and facilitated by ITensor~\cite{itensor-1, itensor-2}. The time-dependent variational principle constrains the time evolution to a specific manifold of matrix product states with a given bond dimension. It projects the action of the Hamiltonian into the tangent space to this manifold and then solves the time-dependent Schrodinger equation solely within the manifold. We use ITensor's GPU backend on a single NVIDIA A100 with 40GB to carry out the MPS-TDVP computation up to $\tau=5$ where we set the time step-size, maximum bond dimension, and cutoff error as $\delta\tau_{\mathrm{TDVP}}=0.25$, $\chi_{\mathrm{max}}=1000$ and $\epsilon_{\mathrm{cutoff}}=10^{-8}$, respectively. Consequently, we have $N_\mathrm{sweeps}=20$ (given by $\tau/\delta\tau_{\mathrm{TDVP}}$) in the MPS-TDVP computation.
\begin{figure}[h!]
	\centerline{\includegraphics[width=10cm]{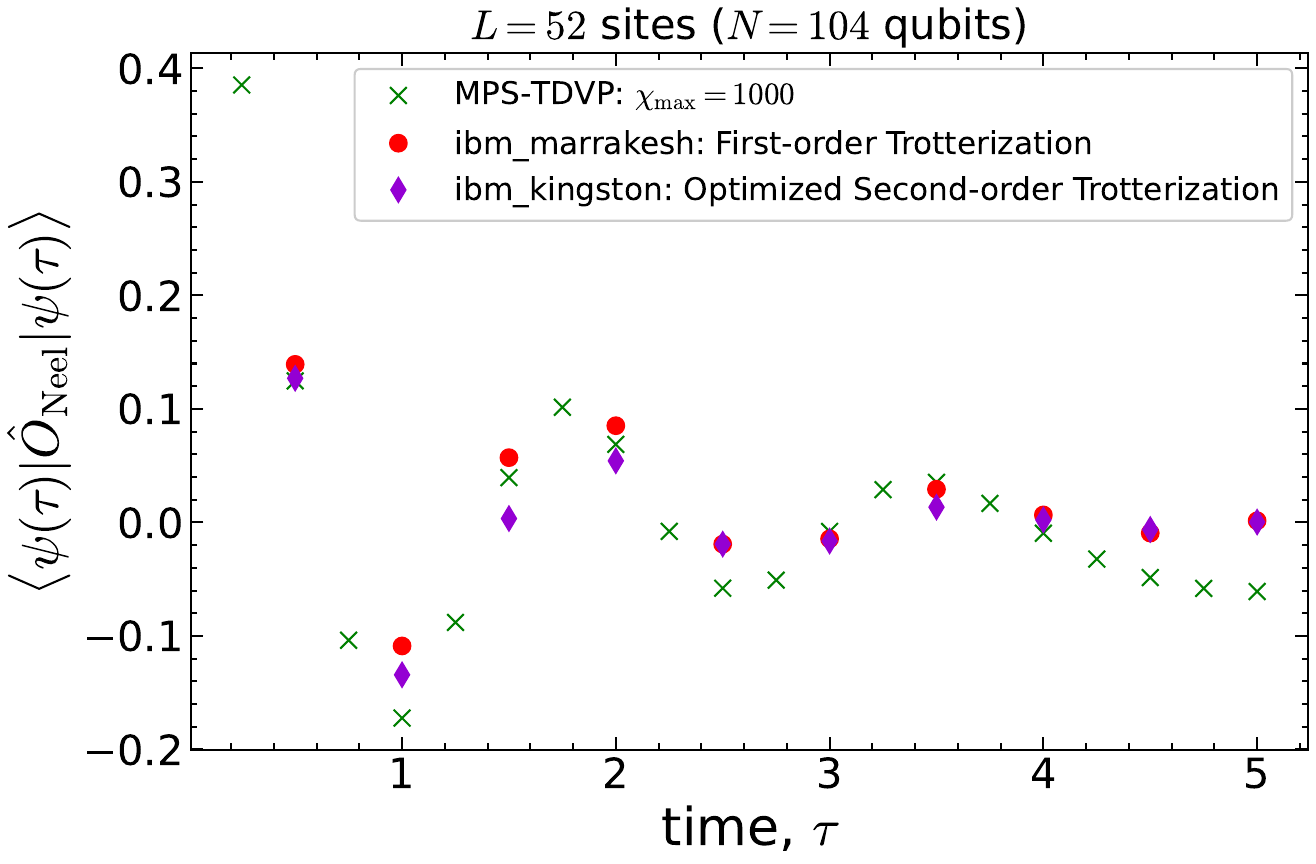}}
	\caption{The time evolution of the N\'eel observable $\langle\psi(\tau)|\hat{O}_{\text{N\'eel}}|\psi(\tau)\rangle$ for $L=52$ sites ($N=104$ qubits) with open boundary conditions where $|\psi(\tau)\rangle = e^{-i H\tau/\hbar}|\psi_{\text{N\'eel}}\rangle$. The Trotter step-size is $\delta\tau=0.5$. Besides, the time step-size, maximum bond dimension, and truncation cutoff in the MPS-TDVP computation are taken as $\delta\tau_{\mathrm{TDVP}}=0.25$, $\chi_{\mathrm{max}}=1000$, and $\epsilon_{\mathrm{cutoff}}=10^{-8}$.}
	\label{fig:Neel-observable-52}
\end{figure}

In Fig.~\ref{fig:Neel-observable-52}, we present the expectation value of the N\'eel observable, computed from the time-evolved state starting from the initial N\'eel state for a system with $L=52$ sites ($N=104$ qubits). We utilized first-order Trotterization and optimized second-order Trotterization circuit implementations on \texttt{ibm$\_$marrakesh} and \texttt{ibm$\_$kingston}, respectively, along with QEM methods to reduce errors in the noisy quantum computers. We observe good agreement between the results obtained from the IBM quantum computers and those from the classical MPS-TDVP method up to a simulation time of $\tau \leq 4$. However, at larger times, such as $\tau > 4$, we encounter issues due to the increasing number of noisy two-qubit controlled-Z (\texttt{cz}) gates, which have a higher median error rate. This problem arises with an increasing number of Trotter steps in the $N=104$ qubit system, as outlined in Table~\ref{tab:FHM-statistics-1st} and Table~\ref{tab:FHM-statistics-2nd}. Consequently, the computations suffer from greater errors that are not efficiently mitigated by the QEM methods.

\begin{figure}[t!]
\centerline{\includegraphics[width=0.48\textwidth]{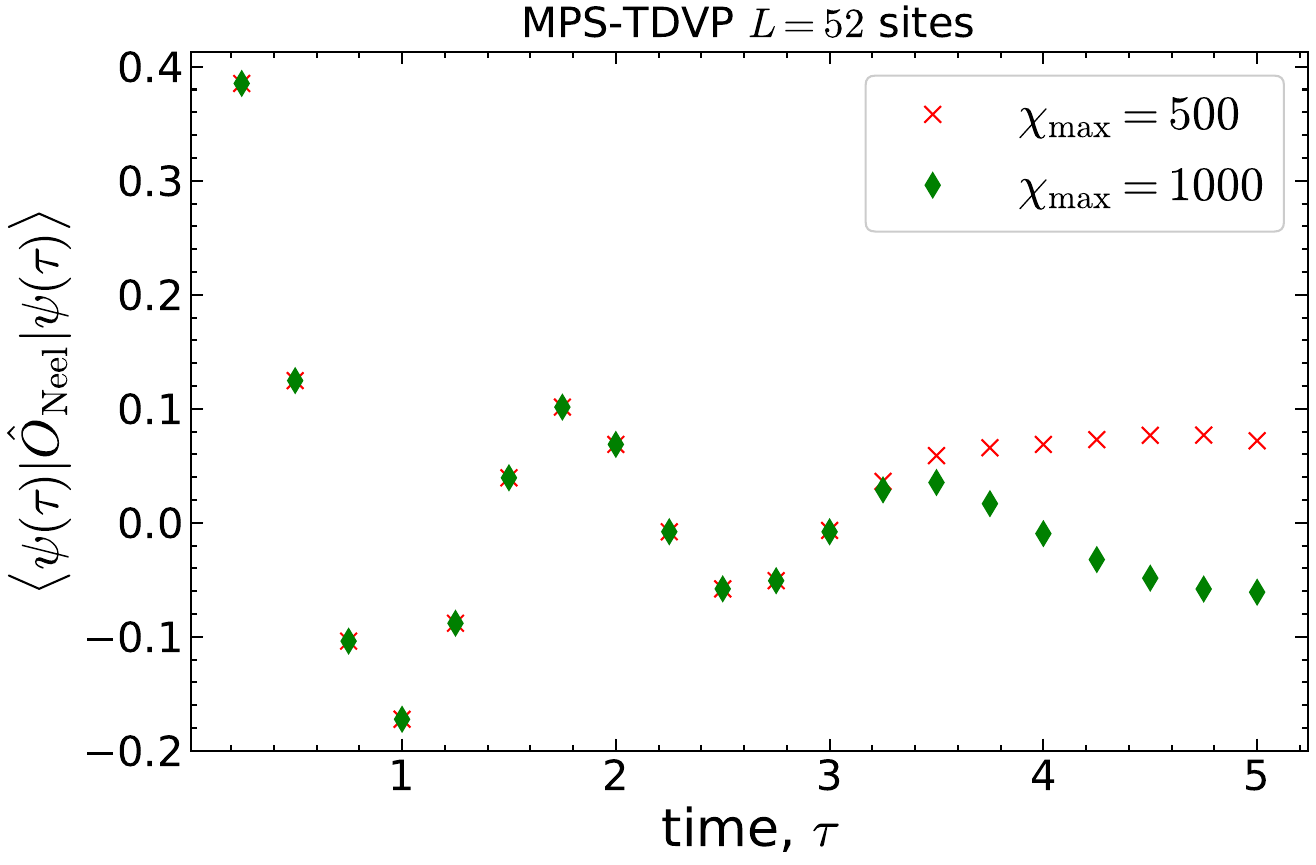}\hspace{2mm}\includegraphics[width=0.48\textwidth]{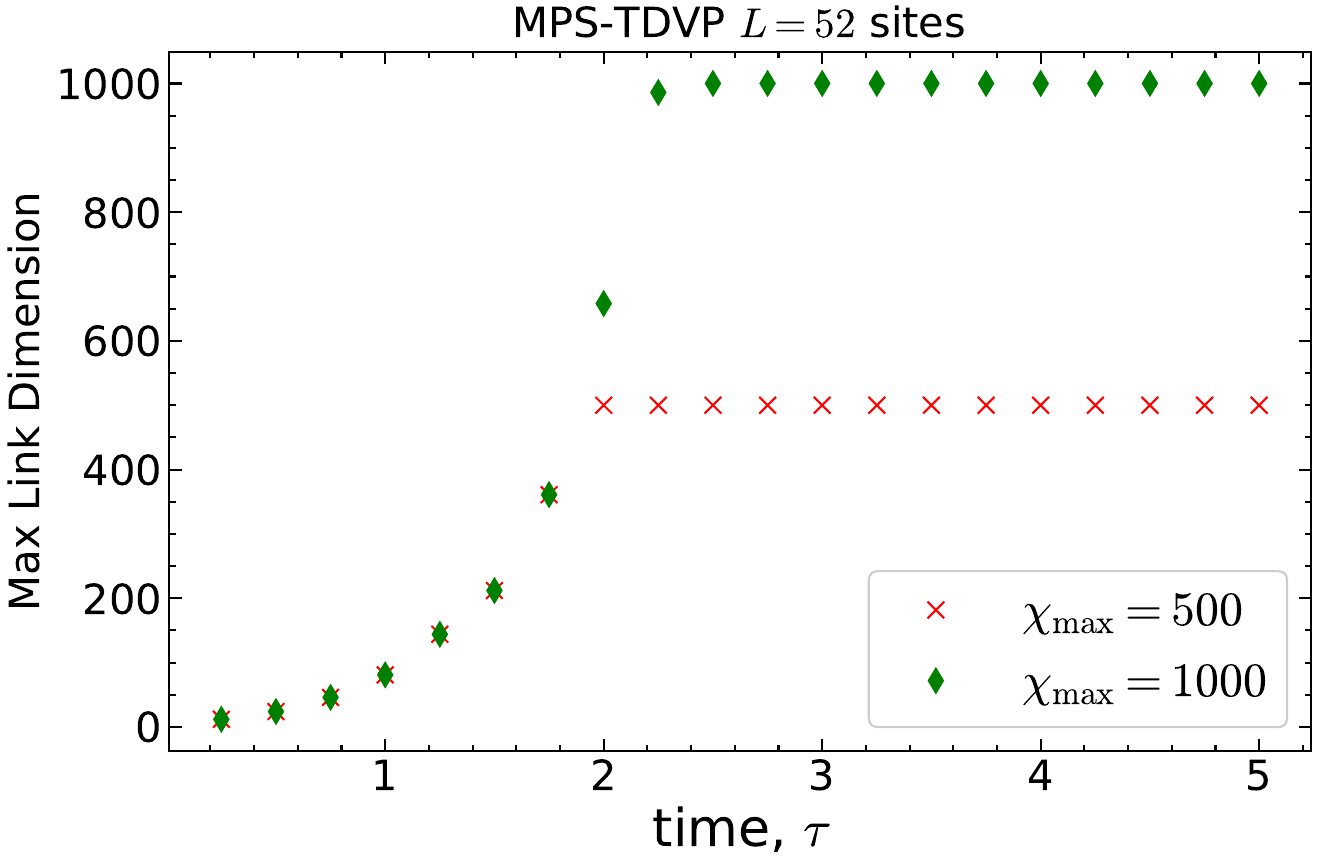}}\vspace{2mm}
\centerline{\includegraphics[width=0.48\textwidth]{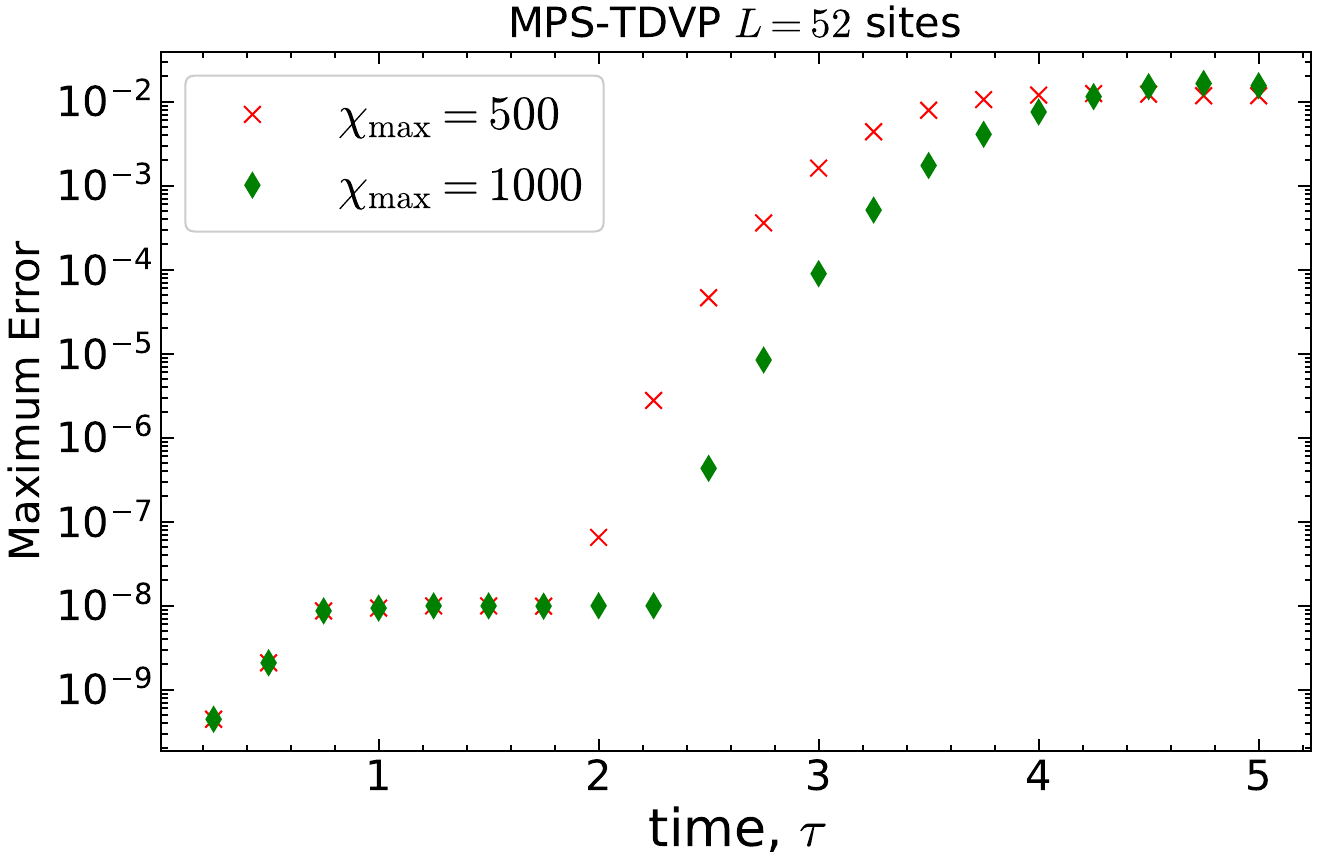}\hspace{2mm}\includegraphics[width=0.48\textwidth]{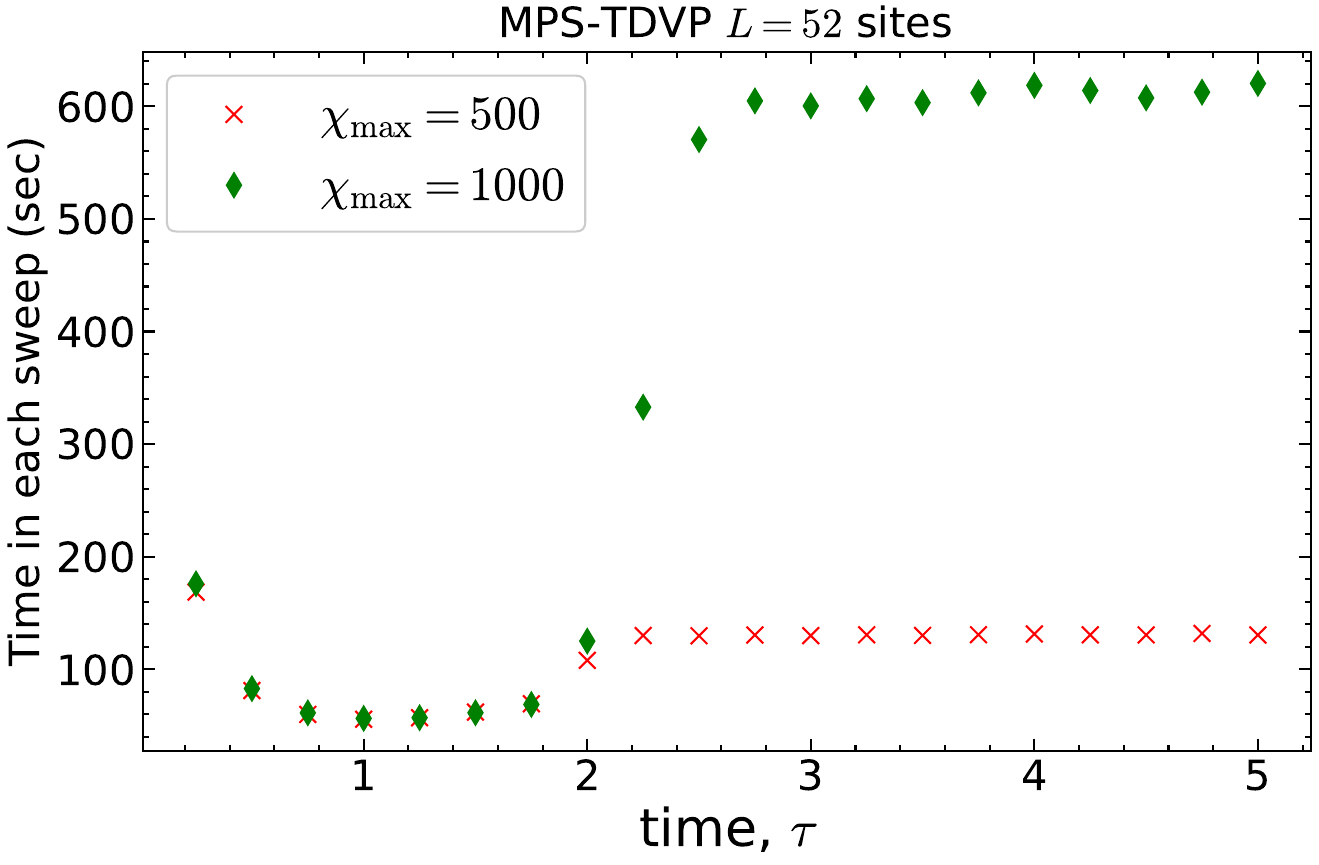}}
\caption{(Upper left figure) The expectation of the N\'eel observable with respect to the time-evolved state during the time evolution using the MPS-TDVP method for the Fermi-Hubbard chain with $L=52$ sites with two different maximum bond dimensions $\chi_{\mathrm{max}}=500$ and $\chi_{\mathrm{max}}=1000$. (Upper right figure) The maximum link dimension associated with the time-evolved MPS during the time evolution in the MPS-TDVP method. (Lower left figure) The maximum truncation error associated with the time-evolved MPS during the time evolution. (Lower right figure) The execution time during each sweep in the MPS-TDVP method with time.}
\label{fig:MPS-details}
\end{figure}

On the other hand, the MPS-TDVP method experiences increasing truncation errors because, at larger times, the time-evolved state demands the MPS with exponentially increasing bond dimensions. This requirement quickly surpasses the maximum bond dimension set within the MPS-TDVP method. In fact, in Fig.~\ref{fig:MPS-details} (upper right figure), we observe that the maximum link dimension representing the maximum dimension of the link in the time-evolved MPS, increases exponentially with time, ultimately getting truncated at the fixed maximum bond dimension $\chi_{\mathrm{max}}$ defined prior to the computation. In addition, the maximum truncation error during a single sweep in the MPS-TDVP method represents the largest of the truncation errors $\epsilon_{\mathrm{truc}}$ associated with discarding the smallest singular values $\lambda_{n}$ at each link in the time-evolved MPS, as follows,
\begin{equation}
    \epsilon_{\mathrm{truc}}=\frac{\sum_{n\in \mathrm{discarded}}\lambda_{n}^{2}}{\sum_{n}\lambda_{n}^{2}}\,,
    \label{eq:MPS-truncation}
\end{equation}
To keep the maximum error within the given cutoff $\epsilon_{\mathrm{cutoff}}$, the maximum bond dimension $\chi_{\mathrm{max}}$ has to be set to a higher value. As we can see in Fig.~\ref{fig:MPS-details} (lower left figure), a smaller $\chi_{\mathrm{max}}$ results in significantly larger truncation errors at longer time intervals. Consequently, as shown in Fig.~\ref{fig:MPS-details} (upper left figure), a smaller maximum bond dimension $\chi_{\mathrm{max}}$ leads to deviations in the computed expectation values compared to those obtained with a larger maximum bond dimension.
Therefore, we do not use the MPS-TDVP results for $\tau > 4$ as validation for the results obtained from the IBM quantum computers. Furthermore, increasing the maximum bond dimension $\chi_{\mathrm{max}} > 1000$ for the fermionic system with $L=52$ sites rapidly exceeds the total memory capacity of the 40GB NVIDIA A100 GPU, failing ITensor's MPS-TDVP computations after more than $N_{\mathrm{sweeps}} > 10$. In retrospect, MPS-based methods can efficiently simulate low-entanglement states, such as ground states, in one-dimensional quantum systems. However, under time evolution, quantum states that are not eigenstates of the Hamiltonian can develop long-range correlations, leading to volume-law scaling of entanglement entropy, meaning that entanglement entropy scales with the size of the corresponding subsystem. Representing such highly entangled states requires a bond dimension that increases exponentially with the system size. Consequently, simulating these evolved states during time evolution, even in large one-dimensional quantum systems, demands prohibitive computational resources. We have already observed this increased computational demand with the MPS-TDVP method while analyzing the Fermi-Hubbard model with over 100 qubits. In contrast, although current noisy quantum computers have limited coherence times that restrict simulation duration, future quantum computers, in conjunction with the scalable quantum simulation circuits developed in this work, will be capable of extending simulation times for larger systems, allowing simulations that surpass the capabilities of classical MPS-based methods and overcoming the entanglement barrier.

Moreover, the average execution (QPU) time for the first-order Trotterization and optimized second-order Trotterization from $\tau=0.5$ (single Trotter step) to $\tau=5$ (ten Trotter steps) on the \texttt{ibm$\_$kingston} system is 14 minutes and 54.8 seconds and 16 minutes and 26.8 seconds, respectively, for $L=10$ sites (20 qubits). On the other hand, for $L=52$ sites ( 104 qubits) it takes 15 minutes and 49 seconds for the first-order Trotterization on \texttt{ibm$\_$marrakesh} and 15 minutes and 40 seconds for the optimized second-order Trotterization on \texttt{ibm$\_$kingston}. In contrast, the execution time during each sweep of the time evolution in the MPS-TDVP computation increases rapidly as time progresses and requires significantly more time for larger values of $\chi_{\mathrm{max}}$. For example, as seen in Fig.~\ref{fig:MPS-details} (lower right figure), in MPS-TDVP computation, after more than 10 sweeps, the bond dimension of the MPS state quickly reaches its maximum limit, with each sweep requiring 600 seconds to complete. Therefore, the IBM quantum computers can more straightforwardly explore larger time scales compared to out-of-the-box packages executing MPS-TDVP on setups like a single GPU, where dedicated multi-CPU and/or multi-GPU configurations are needed to achieve competitive results.

\section{Conclusion and Outlook}\label{sec:conclusion}
In this work, we present a first-order Trotterization for the one-dimensional Fermi-Hubbard model, which is efficient on quantum computers with limited connectivity between qubits, such as superconducting quantum computing platforms. We then extend the first-order implementation to an optimized second-order implementation. 
Our Trotterization requires only the nearest-neighbor and next-nearest-neighbor qubit interactions to implement the hopping terms, interaction terms, and chemical potential terms of the Fermi-Hubbard model.
Based on the proposed Trotterization implementations, we investigate the real-time dynamics in the model using IBM's superconducting quantum computers, which feature over 100 qubits. Specifically, we investigate the relaxation dynamics of the staggered magnetization (or Néel order) in the Fermi-Hubbard model by computing the expectation value of the Néel observable with respect to the time-evolved state. We perform our computations on the quantum devices \texttt{ibm$\_$marrakesh} and \texttt{ibm$\_$kingston} for two cases: one with \( L=10 \) sites (10 qubits) and another with \( L=52 \) sites (104 qubits). Moreover, our first-order and second-order Trotterization circuits are scalable as the corresponding circuit depth associated with a single Trotter step does not depend on the number of qubits. Therefore, in our case, the maximum number of Trotter steps, which sets the total simulation time, is limited only by the total coherence time of the IBM quantum devices. As detailed in Table \ref{tab:FHM-statistics-1st}, for implementation of first-order Trotterization across ten Trotter steps, with a Trotter step size of \(\delta\tau=0.5\), and simulating up to a time \(\tau=5\), the circuit depth and two-qubit gate depth (specifically, the \texttt{cz} gate depth) are 518 and 181, respectively, for a system with \(L=10\) sites (20 qubits) at \texttt{ibm$\_$kingston}. In contrast, for \(L=52\) sites (104 qubits) at \texttt{ibm$\_$marrakesh}, the circuit depth and \texttt{cz} gate depth are 515 and 178, respectively. Additionally, as shown in Table \ref{tab:FHM-statistics-2nd} for the optimized second-order Trotterization implementation, with the same simulation time of \(\tau=5\), the circuit depth and \texttt{cz} gate depth are 840 and 263, respectively, for \(L=10\) sites, whereas, for \(L=52\) sites, the values are 860 and 263, respectively. We observe that for both the first-order and the optimized second-order Trotterization circuits across ten Trotter steps, the circuit depth and \texttt{cz} gate depth remain nearly identical for 20 qubits and 104 qubits, which supports our scalable implementation of the time evolution operator for the one-dimensional Fermi-Hubbard model. Finally, for the basis gate set consists of single-qubit gates  \texttt{rx}, \texttt{rz}, \texttt{x}, \texttt{sx} and a two-qubit gate  \texttt{cz}, the circuit depth approximately scales as $d^{(1)}_{\mathrm{Trotter}} \sim 52\,r $ and $d^{(2)}_{\mathrm{Trotter}} \sim 90\,r$ with the number of Trotter steps $r$, for first-order and optimized second-order Trotterization circuits, respectively, independent of the system size or total number of qubits. On the other hand, for the first-order and optimized second-order Trotterization, the scaling of the total number of two-qubit gate \texttt{cz}, $G^{(i)}_{\texttt{cz}}$ with respect to the total number of Trotter steps $r$ and total number of qubits $N=2L$ are $G^{(1)}_{\texttt{cz}}\sim 4.53\,r^{1.02}N^{1.04}$ and $G^{(2)}_{\texttt{cz}}\sim 7.84\,r^{0.98}N^{1.03}$, respectively, showing polynomial dependence on $r$ and $N$.

We observe excellent agreement between our exact computations and the results obtained from both first-order and optimized second-order Trotterization, combined with quantum error mitigation techniques, using \texttt{ibm$\_$kingston} for the case of the fermionic system with \( L=10 \) sites (20 qubits), which validates our Trotterization circuit implementation of the time evolution operator in the one-dimensional Fermi-Hubbard model. In the case of the larger system with \( L=52 \) sites (104 qubits), we find good consistency between the temporal variation of the expectation value of the Néel observable, calculated using \texttt{ibm$\_$marrakesh} for first-order Trotterization, and using \texttt{ibm$\_$kingston} for optimized second-order Trotterization. These results align well with those obtained through a classical approximation method, referred to here as the MPS-TDVP method, which is based on the matrix product state (MPS) and the time-dependent variational principle (TDVP) and is executed using the ITensor GPU backend on a single 40GB NVIDIA A100 GPU, covering a simulation time up to \( \tau=5 \).

Finally, our work demonstrates the potential of IBM superconducting quantum computers to investigate the properties of large-scale quantum systems. Furthermore, we can expand their applicability to study more intricate scenarios involving the real-time dynamics of the one-dimensional Fermi-Hubbard model on a larger scale, which includes probing the spatially resolved propagation of quantum correlations, the separate dynamics of spin and charge, the dynamics of entanglement entropy, and the scrambling of quantum information. In passing, the Bose-Hubbard model also describes the dynamics of bosons on a lattice, which, similar to the Fermi-Hubbard model, is characterized by the competition between the particle hopping and on-site interaction terms, and the quantum simulation of the Bose-Hubbard model is straightforward by mapping the bosonic creation and annihilation operators onto the qubit operations. However, unlike the Fermi-Hubbard model, which adheres to the Pauli exclusion principle (allowing a maximum of two fermions at a single lattice site), the Bose-Hubbard model permits an arbitrary number of bosons at a single site. This results in a larger-dimensional Hilbert space for each site. Consequently, capturing this higher-dimensional Hilbert space in quantum simulation requires additional qubits, thereby increasing qubit overhead for larger systems. Finally, our work illustrates that IBM quantum computers can explore larger time scales more effectively in scenarios involving large entanglement, as in the case of the Fermi-Hubbard model, compared to standard tensor network methods based on out-of-the-box packages run on single GPUs and necessitate dedicated numerical implementations on multi-CPU or multi-GPU configurations to achieve competitive results for such classical methods.

\section*{Acknowledgment}
\noindent
We thank Subir Sachdev for discussions. VK and VRP are supported by the U.S. Department of Energy, Oﬃce of Science, National Quantum Information Science Research Centers, Co-design Center for Quantum Advantage under Contract No. DE-SC0012704
K.Y. is supported by the Brookhaven National Laboratory LDRD 24-061, 25-033, and the U.S. Department of Energy, Office of Science, Grants No. DE-SC0012704.
This research used quantum computing resources of the Oak Ridge Leadership Computing Facility, which is a DOE Office of Science User Facility supported under Contract DE-AC05-00OR22725.
This research used resources of the National Energy Research Scientific Computing Center (NERSC), a Department of Energy Office of Science User Facility under Contract No. DE-AC02-05CH11231 using NERSC award DDR-ERCAP0034484.
This research used resources of the National Energy Research Scientific Computing Center, a DOE Office of Science User Facility supported by the Office of Science of the U.S. Department of Energy under Contract No. DE-AC02-05CH11231 using NERSC award NERSC DDR-ERCAP0033558. 
IBM, the IBM logo, and ibm.com are trademarks of International Business Machines Corp., registered in many jurisdictions worldwide. Other product and service names might be trademarks of IBM or other companies. The current list of IBM trademarks is available at \url{https://www.ibm.com/legal/copytrade}.

\section*{Data Availability}
The data that supports the findings of this study are available within the article.

\section*{Author Declarations}
The authors have no conflicts to disclose.

\appendix

\section{Quantum Error Mitigation}\label{app:error_mitigations}

One of the key obstacles in executing quantum algorithms on current quantum hardware, such as IBM Quantum processors, is the prevalence of errors and noise. To overcome these limitations, researchers have developed quantum error correction (QEC) strategies. However, implementing quantum error correction (QEC) is not practical for large problems on today's noisy quantum computers because QEC needs a huge number of extra qubits, more than what is currently available, even with new optimizations \cite{Kivlichan-Improved, lee2021even}.
 
In contrast, quantum error mitigation (QEM) accepts the inherent noise and imperfections of near-term quantum devices and applies pre- and post-processing or circuit-level techniques to suppress or reduce their impact on computational outcomes. Unlike QEC, QEM generally requires little to no additional qubit overhead, making it more feasible for current hardware. Over the past few years, a diverse set of QEM strategies, ranging from zero-noise extrapolation to symmetry-based methods, have been developed and successfully demonstrated in practical quantum computing applications \cite{Kim-error-mitigation, kim2023evidence, charles2305simulating, chowdhury2024enhancing, choi2025quantum}.
To mitigate the effects of hardware-induced errors and noise in our experiments, we integrate four complementary QEM techniques, each targeting different stages of the computation: Twirled Readout Error Extinction (TREX) for measurement error symmetrization, Dynamical Decoupling (DD) for mitigating decoherence during idle periods, Pauli Twirling (PT) for converting coherent errors into stochastic noise, and Zero-Noise Extrapolation (ZNE) for estimating noiseless expectation values. Detailed descriptions and implementations of each method are provided in the following subsections~\ref{TREX}-\ref{ZNE}.

\subsection{Twirled Readout Error Extinction}\label{TREX} 

Twirled Readout Error Extinction (TREX) is a straightforward method for mitigating measurement errors in quantum computers \cite{PhysRevA.105.032620}. It works by randomly applying a flip (a Pauli-X gate) to certain qubits right before they're measured. After the measurement, the corresponding classical bits are then flipped back, correcting for the effect of the applied gate. This simple trick helps to reduce the impact of errors that occur during the final readout step.
Although this method has no effect in a perfect, noise-free environment, it has a powerful effect when real-world noise is present. By introducing random flips, the method transforms complicated, state-dependent errors into a simpler, uniform scaling factor, making the errors much easier to correct.
By randomly changing the way it measures, TREX simplifies complex measurement errors, turning them into a much more predictable problem. Instead of being intertwined and unpredictable, each qubit's error becomes independent, affecting only its own result. This makes the errors easy to correct. One of the biggest advantages is that TREX doesn't need a detailed understanding of the specific noise in the system, which makes it a robust solution even when the noise is difficult to figure out. However, it is not as effective at fixing errors that are strongly connected between multiple qubits.
We use $10$ samples for the TREX process in our experiments.

\subsection{Dynamical Decoupling}\label{DD}

Dynamical Decoupling (DD) is a quantum error mitigation technique aimed at suppressing decoherence and crosstalk effects, particularly those induced by interactions with spectator qubits. It achieves this by applying carefully designed periodic sequences of fast control pulses that average out system-environment couplings, thereby effectively canceling their contribution to the error dynamics \cite{viola1999dynamical}. In practice, DD applies sequences of single-qubit rotations to otherwise idle qubits, effectively toggling their basis and averaging out noise contributions from environmental interactions and neighboring qubits. As a result, the effective coherence time of the circuit is extended.
The effectiveness of DD has been verified through practical tests in various environments \cite{ezzell2022dynamical, niu2022effects, Kim-error-mitigation, kim2023evidence, charles2305simulating}.
In these experiments, we use ($t/4$, $X$, $t/2$, $X$, $t/4$) sequence in every idling period for the DD implementation, where $X$ represents the $\texttt{XGate}$ and $t$ is the idling time except for the two $\texttt{XGate}$ pulse durations.

\subsection{Pauli Twirling}\label{PT}

\begin{figure}[h!]
\[
\Qcircuit @C=1.2em @R=1.2em {
  & \multigate{1}{~~~\texttt{cz}~~~} & \qw & & & & \gate{~1~} & \multigate{1}{~~~\texttt{cz}~~~} & \gate{~3~} & \qw \\
  & \ghost{~~~\texttt{cz}~~~} & \qw &  & \raisebox{0.8cm}{~~=~~~~~} & & \gate{~2~} & \ghost{~~~\texttt{cz}~~~} & \gate{~4~}  & \qw
}
\]
\vspace*{1mm}
\caption{A circuit diagram for Pauli twirling for $\texttt{cz}$ gate. At positions 1, 2, 3, and 4, Pauli gates, $\{ I, \sigma^x, \sigma^y, \sigma^z \}$ are placed.} 
\label{fig:pauli_twriling}
\end{figure}
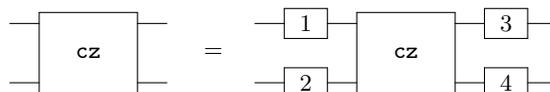

Pauli twirling is a method that converts structured, coherent errors in a quantum circuit into random, noise-like errors. It does this by essentially scrambling the errors. Averaging over a specific set of operations eliminates the predictable, off-diagonal parts of the error in the Pauli basis, $\{ I, \sigma^x, \sigma^y, \sigma^z \}$ and leaves only a form of random noise, making the errors easier to handle and mitigate \cite{bennett1996purification, wallman2016noise, cai2019constructing}.
In PT, each Clifford gate is conjugated by Pauli operators applied before and after the gate, resulting in a transformation that is algebraically equivalent to the original Clifford. The efficiency of this approach has been demonstrated in prior studies \cite{Kim-error-mitigation, kim2023evidence, chowdhury2024enhancing, chowdhury2024capturing, CHOWDHURY2025108526}.
Figure~\ref{fig:pauli_twriling} illustrates the PT method used in this study. We apply this technique specifically to the $\texttt{cz}$ gates, which function as the counterpart of $\texttt{CX}$ gates in our implementation.
The method searches all the combinations of Pauli gates that are mathematically identical up to the global phase, with only the Clifford gate, $\texttt{cz}$.
Since the Pauli gate set has four elements and we have four positions, the search domain is $256 (=4^4)$ cases.
The trivial case is placing the identity gate in the position, $1, 2, 3$, and $4$ in Fig. \ref{fig:pauli_twriling}.
One non-trivial case is putting $\sigma^z, \sigma^x, \sigma^z,$ and $\sigma^x$ at $1, 2, 3$, and $4$, respectively.
We generated ten instances of the base quantum circuit, each incorporating a randomly selected Pauli-twirling gate sequence from the predefined set, which was then applied to the corresponding Clifford gate as outlined in Fig. \ref{fig:pauli_twriling}.

\subsection{Zero-Noise Extrapolation}\label{ZNE}


Zero-Noise Extrapolation (ZNE) is an error mitigation technique that estimates the noise-free expectation value by extrapolating from measurement outcomes obtained at systematically varied noise levels. By analyzing how the results change as the noise increases, it's possible to project what the result would have been if there were no noise at all \cite{Temme-error-mitigation, Li-error-mitigation, giurgica2020digital}.
In our experimental setup, we employ local unitary gate folding \cite{giurgica2020digital} as the noise-scaling mechanism for Zero-Noise Extrapolation. This procedure is selectively applied to two-qubit $\texttt{cz}$ gates, as their error rates are empirically observed to exceed those of single-qubit gates by more than an order of magnitude. To realize noise amplification, we fold the $\texttt{cz}$ operations with scaling factors of 1 (noiseless baseline), 3 (single fold), and 5 (double fold), thereby generating circuits with systematically increased effective noise while preserving the ideal logical action of the original circuit.
ZNE operates under the assumption that noise can be systematically amplified in a controlled fashion, even though the corresponding extrapolation curve may exhibit nonlinearity. In practice, however, quantum noise is frequently non-Markovian and time-dependent, such that noise scaling does not necessarily yield predictable modifications to the circuit output. Consequently, when the underlying noise model is inaccurate or unstable, the extrapolation procedure may produce unreliable or even misleading estimates. When the exact expectation value is known, it is possible to construct a precise fitting curve for extrapolation. This curve, obtained from small-scale estimations, can be extended to larger systems under the assumption that the noise model remains consistent during scaling. However, this assumption frequently fails as the number of qubits increases.

\bibliography{references}
\end{document}